\documentclass[reqno,11pt]{amsart}
\usepackage[utf8]{inputenc}
\usepackage{graphicx}
\usepackage{slashed}
\usepackage{amscd}
\usepackage{tensor}
\usepackage{amssymb}
\usepackage{esint}
\usepackage{enumitem}
\usepackage{accents}
\usepackage[usenames,dvipsnames]{pstricks}
\usepackage{pst-grad} 
\usepackage{pst-plot} 
\usepackage[mathscr]{eucal}
\numberwithin{equation}{section}
\allowdisplaybreaks[1]

\title[The Regularized Hadamard Expansion]{The Regularized Hadamard Expansion}

\author[F.\ Finster]{Felix Finster}
\address{Fakult\"at f\"ur Mathematik \\ Universit\"at Regensburg \\ D-93040 Regensburg \\ Germany}
\email{finster@ur.de}

\author[M.\ Kraus]{Margarita Kraus \\ \\ June 2020}
\address{Fachbereich 08 \\ Johannes-Gutenberg-Univer\-si\-t\"at Mainz \\
D-55099 Mainz \\ Germany }
\email{mkraus@mathematik.uni-mainz.de}

\newtheorem{Def}{Definition}[section]
\newtheorem{Thm}[Def]{Theorem}
\newtheorem{Prp}[Def]{Proposition}
\newtheorem{Lemma}[Def]{Lemma}

\newtheorem{Example}[Def]{Example}

\newcommand{\Thanks}{\vspace*{.5em} \noindent \thanks}
\newcommand{\beq}{\begin{equation}}
\newcommand{\eeq}{\end{equation}}
\newcommand{\Proof}{\begin{proof}}
\newcommand{\QED}{\end{proof} \noindent}
\newcommand{\QEDrem}{\ \hfill $\Diamond$}

\newcommand{\la}{\langle}
\newcommand{\ra}{\rangle}

\newcommand{\R}{\mathbb{R}}

\newcommand{\Z}{\mathbb{Z}}
\newcommand{\N}{\mathbb{N}}

\renewcommand{\O}{{\mathscr{O}}}

\newcommand{\Dir}{{\mathcal{D}}}

\newcommand{\scrL}{\mycal L}
\newcommand{\scrM}{M}
\newcommand{\scrN}{\mycal N}
\newcommand{\tscrM}{\,\tilde{\!\scrM\,}\!}
\newcommand{\tscrN}{\,\tilde{\!\scrN\,}\!}

\DeclareFontFamily{OT1}{rsfso}{}
\DeclareFontShape{OT1}{rsfso}{m}{n}{ <-7> rsfso5 <7-10> rsfso7 <10-> rsfso10}{}
\DeclareMathAlphabet{\mycal}{OT1}{rsfso}{m}{n}
\DeclareMathOperator{\vleck}{\mathcal{V}}

\newcommand{\s}{\mathfrak{s}}

\newcommand{\bitem}{\begin{itemize}[leftmargin=2em]}
\newcommand{\eitem}{\end{itemize}}
\newcommand{\f}{f}
\newcommand{\fa}{a}
\newcommand{\fb}{b}

\begin{document}

\maketitle

\begin{abstract}
A local expansion is proposed for two-point distributions involving an ultraviolet regularization in
a four-dimensional globally hyperbolic space-time.
The regularization is described by an infinite number of functions
which can be computed iteratively by solving transport equations along null geodesics.
We show that the Cauchy evolution preserves the regularized Hadamard structure.
The resulting regularized Hadamard expansion gives detailed and explicit information on the
global dynamics of the regularization effects.
\end{abstract}

\tableofcontents

\section{Introduction}
Hadamard states play an important role in quantum field theory because
they are a suitable starting point for the perturbative treatment
(see for example~\cite{fewster2013necessity, dappiaggiDirac}
or the recent text book~\cite{rejzner}).
From the point of view of microlocal analysis,
quasifree Hadamard states are characterized by the singularity structure
of their two-point distribution as expressed in terms of its wave front set
(see~\cite{radzikowski} or~\cite{hormanderI, strohmaierML}).
Alternatively, this singularity structure becomes apparent when writing the
two-point distribution in local coordinates (see~\cite{waldQFT}).
For a scalar wave or a Klein-Gordon field
on a four-dimensional globally hyperbolic Lorentzian manifold~$(\scrM, g)$,
the two-point distribution~$T(x,y)$ of a Hadamard state
admits locally a {\em{Hadamard expansion}}
of the form (see for example~\cite{hadamardoriginal, waldQFT})
\beq \label{Tform}
T(x,y)  =  \lim_{\varepsilon \searrow 0} \Big( \frac {U(x,y)}{\Gamma_\varepsilon(x,y)} + V(x,y)\:
\log \Gamma_\varepsilon(x,y)+ W(x,y)  \Big)
\eeq
with
\[ \Gamma_\varepsilon(x,y) :=\Gamma(x,y) - i \varepsilon \,\big(\mathfrak{t}(y)-\mathfrak{t}(x) \big) \:, \]
where~$\mathfrak{t}$ is a time function and~$\Gamma(x,y)$ is the geodesic distance squared, with the sign convention that~$\Gamma$ is positive in timelike and negative in spacelike directions.
Moreover, the functions~$U$, $V$ and $W$ should have an expansion in powers of~$\Gamma$
\beq \label{UVWexpand}
U = \sum_{n=0}^{\infty} U_n \:\Gamma^n\:,  \qquad
V = \sum_{n=0}^{\infty} V_n \:\Gamma^n \:, \qquad
W =\sum_{n=0}^{\infty} W_n\:\Gamma^n
\eeq
with smooth coefficient functions~$U_n$, $V_n$ and $W_n$.
The method to obtain the Hadamard expansion is to insert the above ansatz
into the hyperbolic partial differential equation, and to evaluate the resulting
terms order by order in powers of~$\Gamma$. This gives rise to a system
of transport equations along null geodesics which can be solved iteratively. This
method  of {\em{integration along characteristics}} goes back
to Jacques Hadamard and Marcel Riesz~\cite{hadamardoriginal, riesz}; see also
the textbooks~\cite{friedlander1, baer+ginoux}.
Before going on, for clarity we point out that the infinite sums in~\eqref{UVWexpand}
are to be understood as formal power series defined only via the approximation of the
partial sums. This is unproblematic because
the existence of the bi-distribution~$T(x,y)$ in~\eqref{Tform} is guaranteed
using non-perturbative constructions for hyperbolic partial differential equations.
Therefore, the purpose of the Hadamard expansion merely is to describe the
local behavior of the singularity on the light cone.

The Hadamard expansion can be carried out similarly for other linear hyperbolic equations.
In particular, for a Dirac field one chooses a local chart on the Lorentzian
manifold~$(\scrM, g)$ and a local trivialization of the spinor bundle~$S\scrM$.
Then a distributional bi-solution~$P(x,y)$ of the Dirac equation
\[ (\Dir - m) P(x,y) = 0 \]
is of Hadamard form if
(see~\cite{sahlmann2001microlocal} or~\cite[page~156]{hack})
\[ P(x,y) = \lim_{\varepsilon \searrow 0} \; \Dir_x \left( \frac{U(x,y)}{\Gamma_\varepsilon(x,y)}
+ V(x,y)\: \log \Gamma_\varepsilon(x,y) + W(x,y) \right) , \]
where~$U$, $V$ and~$W$ 
are now mappings between the corresponding spinor spaces,
\beq \label{UVWbundle}
U(x,y),\; V(x,y),\; W(x,y) \;:\; S_y\scrM \rightarrow S_x\scrM \:.
\eeq

The motivation and interest in regularizing the Hadamard expansion
evolved as follows. Generally speaking, ultraviolet regularizations are an important tool
in quantum field theory needed in order to make divergent expressions mathematically
well-defined. In the simplest case of a scalar field in Minkowski space, 
the most obvious regularization is obtained by not taking the limit~$\varepsilon \searrow 0$
in~\eqref{Tform}, i.e.\ by setting
\begin{gather}
T^\varepsilon(x,y)  =  \frac {U(x,y)}{\Gamma_\varepsilon(x,y)} + V(x,y)\:
\log \Gamma_\varepsilon(x,y)+ W(x,y) \label{Teps} \\
\text{with} \qquad \Gamma_\varepsilon(x,y) = (y-x)^2 - i \varepsilon \,\big(y^0 - x^0 \big) \label{xi2eps}
\end{gather}
(here~$(y-x)^2 = \sum_{i,j=0}^3 g_{ij} (y-x)^i (y-x)^j$ is the Minkowski inner product
with sign convention~$g_{ij} = \text{diag}(1,-1,-1,-1)$).
Then the parameter~$\varepsilon>0$ gives the length scale of the regularization; it can
be thought of as a microscopic length scale (for example the Planck scale).
Of course, the above choice for~$T^\varepsilon$ is very special,
because there are many other ways to regularize.
In the renormalization program, one shows that the limit~$\varepsilon \searrow 0$
of all observables exists if the masses and coupling constants are suitably rescaled
in powers of~$\varepsilon$. It is generally believed (although in most situations not proven)
that this limit should not depend on how the regularization is chosen.
With this in mind, in the renormalization program the regularization is merely a mathematical tool
with no physical significance.

This picture changed when physicists began taking the concept seriously
that on a microscopic length scale (often thought of as the
Planck scale) space-time should no longer be modeled by Minkowski space
or by a Lorentzian manifold, but that it could have instead a different, yet unknown, microstructure.
One approach is string theory, where the fundamental objects are strings propagating
in a higher-dimensional space-time (see for example~\cite{becker+schwarz}).
Another approach is non-commutative geometry,
where the microstructure is described by generalizing the commutative algebra of
functions on a manifold to a non-commutative algebra
(see~\cite{connes} or the generalization to Lorentzian signature in~\cite{besnard}).
Noncommutative geometry gives rise to an interesting connection between ultraviolet and infrared
properties of the resulting quantum field theories~\cite{grosse+wulkenhaar}.
In other approaches like causal dynamical triangulations and causal sets
(see for example~\cite{loll, sorkin}), the microstructure of space-time is discrete.
In all these approaches, which are sometimes subsumed under the keyword {\em{``quantum geometry,''}}
the microscopic structures can be regarded as a ``regularization'' of the 
usual space-time continuum for very small distances. In other words, in these approaches
the regularization does have a physical significance, because it models the
microscopic structure of space-time.
These concepts are also implemented in the theory of causal fermion systems,
being a recent approach to fundamental physics (see~\cite{cfs}
or the survey article~\cite{dice2014}). In this theory, the {\em{regularized}} objects
are considered as the physical objects. In particular, the regularization encodes
physical parameters like the bosonic masses and the coupling constants. Consequently,
if the properties of the regularization vary in space-time, this should lead to
a dynamical behavior of these physical parameters.
This idea was first explored in~\cite{dgc} for the gravitational coupling constant.

In view of these developments, it is an important task to describe the regularization
effects and to study their dynamical behavior.
The present paper is the first work dedicated to the systematic study of this problem.
Taking a conservative approach, we do not modify the underlying equations
(for regularizations which do modify the underlying equations see for example~\cite[Chapter~4]{pfp}).
Thus the regularized scalar distribution~$T^\varepsilon(x,y)$ should still satisfy the Klein-Gordon equation
\beq \label{KGeps}
\big( \Box + m^2 \big) \,T^\varepsilon(x,y) = 0
\eeq
(where~$\Box = \nabla^j \nabla_j$ is the scalar wave operator, where we
always choose the signature convention~$(+,-,-,-)$).
This implies that~$T^\varepsilon$, and therefore also the dynamics of the regularization,
is determined from initial data on a Cauchy surface.
This observation by itself is not very helpful because, in general, the Cauchy problem cannot be solved explicitly.
The interesting point, which also makes the connection to the Hadamard expansion~\eqref{Tform},
is that many {\em{regularization effects}} are also {\em{described by transport equations}}
along null geodesics. Compared to~\eqref{Tform}, the resulting regularized Hadamard expansion contains
a lot of additional explicit information on the dynamics of the regularization effects.

We now explain our results and put them in the context of previous works.
The first systematic study of regularization effects was carried out in~\cite[Chapter~4]{pfp}
in Minkowski space, however in a more general setting where the underlying equations
were modified by the regularization.
This analysis distinguished certain regularization effects 
(arising from the so-called mass expansion) as being {\em{universal}} in the
sense that they do not depend on the details of the regularization.
The resulting contributions to~$T^\varepsilon(x,y)$ are used in a crucial way in the 
analysis of the continuum limit (see also the recent textbook~\cite{cfs}).
It is precisely these contributions which are described by transport
equations and which will be specified in detail below.
The analysis in~\cite[Chapter~4]{pfp} also revealed that other regularization effects 
(in particular those which describe the behavior near the diagonal~$x=y$)
are {\em{not}} universal and seem to depend on the details of the regularization.
In our context, these contributions to~$T^\varepsilon$ can be characterized by the fact that
they cannot be described by transport equations.

The just-described distinction between contributions which can and cannot be described
by transport equations can be made precise
in terms of a scaling behavior in~$\varepsilon$, as we now explain.
For clarity, we begin in Minkowski space. 
We write Minkowski vectors as~$x=(x^0, \vec{x})$ with~$x^0 \in \R$ and~$\vec{x} \in \R^3$.
Then a simple regularization is obtained by
analytic continuation of the time variable to the complex plane via the replacement
(for details see~\cite[\S2.4.1]{cfs} or Example~\ref{exexplicit} below)
\beq \label{ieps}
(y-x)^0 \rightarrow (y-x)^0 - i \varepsilon \:.
\eeq
Regularizing~$\Gamma$ in this way,
\beq \label{Gammareg}
\Gamma \rightarrow (y^0 - x^0 - i \varepsilon)^2 - \big| \vec{y} - \vec{x} \big|^2
= \Gamma - 2 i \varepsilon \, \big(y^0 - x^0\big) - \varepsilon^2 \:,
\eeq
one sees that this so-called {\em{$i \varepsilon$-regularization}} is indeed very similar to the
procedure in~\eqref{Teps}. Namely, it differs from~\eqref{xi2eps} only by a factor of two
(which is irrelevant because it can be absorbed into~$\varepsilon$) and the term of the order~$\varepsilon^2$.
The replacement~\eqref{ieps} has the advantage that it preserves the Klein-Gordon equation~\eqref{KGeps}
without error terms.
Generally speaking, we want to neglect terms which are small even near the light cone.
More precisely, by ``near the light cone''  we mean that the spatial distance of $y-x$ from the light cone
is of the order~$\varepsilon$ of the regularization scale. Thus, setting~$\xi=y-x$,
\[ \Big| \big| \xi^0 \big| - \big| \vec{\xi} \,\big| \Big| \lesssim \varepsilon \:. \]
On this scale,
\[ |\Gamma| \lesssim \varepsilon\, \big| \xi^0 \big| \:, \]
showing that the summand~$2 i \varepsilon \xi^0$ in~\eqref{Gammareg} 
is {\em{not}} small and cannot be neglected. But the term quadratic in~$\varepsilon$ in~\eqref{Gammareg}
can be omitted if we write the error term as
\[ \varepsilon^2 = \varepsilon\, \big| \xi^0 \big| \; \O\bigg(\frac{\varepsilon}{\big| \xi^0 \big|} \bigg) 
\qquad \text{or alternatively} \qquad
\varepsilon^2 = \Gamma\: \O\bigg(\frac{\varepsilon^2}{\Gamma} \bigg) \:. \]
This simple observation, which was indeed the starting point for developing the formalism of the continuum limit,
extends to a much more general setting. Namely, it leads to the general rule
to take into account only the terms linear in~$\varepsilon$, but to neglect quadratic and higher orders.
This rule was derived and explained in different variants in~\cite[Chapter~4]{pfp}
and~\cite[Section~2.4]{cfs}. 
It can be understood already in Minkowski space without interaction
from the fact that only the structure of the linear terms in~$\varepsilon$ is robust when
considering general regularizations (as is explained in~\cite[\S2.4.2]{cfs} in the
example of linear combinations of $i \varepsilon$-regularizations; for the general
analysis see~\cite[Chapter~4]{pfp}).
If an interaction is present, one sees explicitly from the analysis
in~\cite[Appendix~D]{pfp} and~\cite[Appendix~F]{cfs}
that the quadratic terms in~$\varepsilon$ cannot be described
by transport equations. With this in mind, in this paper we always restrict attention to
the terms linear in~$\varepsilon$. Writing the resulting error term in the form
\beq \label{fehler}
\bigg(1 + \O\Big( \frac{\varepsilon^2}{\Gamma} \Big) \bigg) \:,
\eeq
this error term is dimensionless and can be used in curved space-time as well
(we remark that these error terms will be refined in Section~\ref{sectransport}; see~\eqref{fehlerneu}).

We now give an outline of our main results. In Section~\ref{secscal}
a Klein-Gordon field is considered. For the regularized Hadamard expansion
of a bi-solution~$T^\varepsilon$ we make the ansatz
\beq \begin{split}
&T^\varepsilon(x,y)  = \frac{X_{-1}(x,y)}{\Gamma_{[-1]}(x,y)} \\
& + \sum_{n=0}^{\infty} X_n(x,y)\; \Gamma_{[n]}(x,y)^n\:\log \Gamma_{[n]}(x,y)
+ \sum_{n=0}^{\infty} Y_n(x,y)\: \Gamma_{\{n\}}(x,y)^n \:,
\end{split} \label{reghadamard}
\eeq
where
\begin{align}
\Gamma_\bullet(x,y) &= \Gamma(x,y) + i \varepsilon \,\f_\bullet(x,y) \label{Gammaf} \\
X_n(x,y) &= A_n(x,y) + i \varepsilon \,\fa_{n}(x,y) \label{XA} \\
Y_n(x,y) &= B_n(x,y) + i \varepsilon \,\fb_{n}(x,y) \label{YB}
\end{align}
with smooth real-valued functions~$A_{-1}$, $A_n$ and~$B_n$ as well as
real-valued continuous functions~$\f_\bullet$, $\fa_\bullet$ and~$\fb_\bullet$ which are smooth away from the diagonal~$x=y$
(where the bullets stand for any subscripts).
Going to Minkowski space and choosing all the functions~$\f_\bullet$ equal to~$x^0-y^0$,
we recover the $i \varepsilon$-regularization~\eqref{Teps}. Moreover,
taking the limit~$\varepsilon \searrow 0$, we recover~\eqref{Tform}, provided that
the functions~$\f_\bullet(x,y)$ have the same sign as~$\mathfrak{t}(x) - \mathfrak{t}(y)$.

The functions~$A_{-1}$, $A_n$ and~$B_n$ are shown to satisfy the transport
equations of the standard Hadamard expansion (as worked out for example in~\cite[Section~2]{baer+ginoux}).
The regularization is described by the sequences of functions
\[ \f_{[-1]}, \;\f_{[0]}, \: \f_{[1]}, \cdots \qquad \text{and} \qquad
\f_{\{0\}}, \: \f_{\{1\}}, \cdots \]
as well as
\[ \fa_{-1}, \;\fa_{0}, \: \fa_{1}, \cdots \qquad \text{and} \qquad
\fb_{0}, \: \fb_{1}, \cdots \:. \]
Our main result is to show that these functions also satisfy transport equations,
which we derive and analyze. Before stating our main results, we
need to briefly explain the freedom in modifying the above expansion
while preserving the Klein-Gordon equation~\eqref{KGeps}.
First, we can obviously
\beq \label{free1}
\text{multiply~$T^\varepsilon$ by a complex constant}\:.
\eeq
Moreover, we can add to~$T^\varepsilon$ any smooth solution of the Klein-Gordon equation.
This amounts to the freedom of
\beq \label{free2}
\text{choosing~$B_0$ and~$\fb_{0}$ arbitrarily}\:.
\eeq
Having already chosen the standard Hadamard coefficients, the above freedom~\eqref{free1} and~\eqref{free2}
reduces to
\begin{gather}
\text{multiply~$T^\varepsilon$ by an imaginary constant~$\sim \varepsilon$} \label{free3} \\
\text{choose~$\fb_{0}$ arbitrarily}\:. \label{free4}
\end{gather}
We first state our local result.
\begin{Thm} \label{thmmain}
The functions~$\f_\bullet(x,y)$, $a_\bullet(x,y)$ and~$b_\bullet(x,y)$ must satisfy transport equations along
the light cone (for details see Proposition~\ref{prptransport2} below). These equations can be solved iteratively to obtain 
continuous solutions which are smooth away from the diagonal~$x=y$.
The solution of the transport equations for~$\f_{[-1]}$ involves one free parameter
along each null geodesic.
The solutions of all the other transport equations are unique up to the freedom~\eqref{free3}
and~\eqref{free4}. Moreover, the functions~$\f_\bullet$ in~\eqref{Gammaf} 
are antisymmetric in~$x$ and~$y$ in the sense that
the equations
\beq
\f_\bullet(x,y) = -\f_\bullet(y,x) \label{fantisymm} 
\eeq
hold on the light cone.
\end{Thm} \noindent
In order to clarify the global structure of our expansion, we show that the
functions~$\f_\bullet$ are uniquely determined globally by specifying the multiplicative parametrization
of each null geodesic (for details see the family~$\scrL$ in Definition~\ref{defL}).
Having made this choice, the regularized Hadamard expansion is unique, 
up to the obvious freedom~\eqref{free3} and~\eqref{free4} as well as
well-defined error terms.
Estimating these error terms in the time evolution, we prove that
the regularized Hadamard form is preserved by the time evolution
(see Theorem~\ref{thmpropagate}). This theorem can be applied to
prove the existence of the regularized Hadamard expansion in generic space-times (see Theorem~\ref{thmglue}),
simply by adapting glueing constructions which are commonly used for the construction of Hadamard
states in generic space-times.

In Section~\ref{secdirac} our methods and results are extended to the Dirac field.
The regularized Hadamard expansion reads
\beq \label{Pexpand}
P^\varepsilon(x,y) = (\Dir + m)\, T^\varepsilon(x,y) \:,
\eeq
where~$T^\varepsilon(x,y)$ is again of the form~\eqref{reghadamard}.
The transport equations also remain valid if one only keeps in mind that,
similar to~\eqref{UVWbundle}, the functions~$X_\bullet$ and~$Y_\bullet$ 
are now mappings between corresponding fibres of the spinor bundle.

We close with a remark on the symmetry properties of~$T^\varepsilon$ und $P^\varepsilon$
in its two arguments.
We point out that all our results apply without any symmetry assumptions.
The symmetry property~\eqref{fantisymm} is a mere consequence of the assumption
that in the limit~$\varepsilon \searrow 0$, our expansion should go over to the usual
unregularized Hadamard expansion.
Having bi-solutions, the symmetric and anti-symmetric components defined by
\[ \frac{1}{2} \big( T^\varepsilon(x,y)  \pm \overline{T^\varepsilon(y,x)} \big) \qquad \text{and} \qquad
\frac{1}{2} \big( P^\varepsilon(x,y) \pm P^\varepsilon(y,x)^* \big) \]
(where the star denotes the adjoint with respect to the inner product on the spinor spaces)
are again bi-solutions. Therefore, we can perform the regularized Hadamard expansion 
of each of them separately, giving rise to obvious symmetry properties
of the regularized Hadamard coefficients. In particular, for symmetric bi-solutions,
the function~$\fa_{-1}$ has the symmetry property
\[ \fa_{-1}(x,x) = 0 \:, \]
which fixes the arbitrariness~\eqref{free3}. This makes the local
expansion unique up to the usual freedom to add smooth bi-solutions.

The paper is organized as follows. In Section~\ref{secscal} the transport equations
are derived and solved for a Klein-Gordon field. Moreover, we give the proof
of Theorem~\ref{thmmain}.
In Section~\ref{secprop} we work out the global character of the expansion
and prove a theorem on the propagation of regularized singularities
(see Theorem~\ref{thmpropagate}).
In Section~\ref{secex} an explicit example of a regularized Hadamard expansion
is given, and it is used to show the existence of
regularized Hadamard bi-solutions in generic space-times.
In Section~\ref{secdirac} our methods and results are extended to the Dirac equation.
Appendix~\ref{appendixA} contains the detailed computations needed for the
derivation of the transport equations. 

\section{A Klein-Gordon Field} \label{secscal}
Let~$(\scrM, g)$ be a four-dimensional globally hyperbolic Lorentzian manifold with Cauchy time function~$\mathfrak{t}$.
In this section we always restrict attention to a geodesically convex set~$\Omega \subset \scrM$
(see~\cite[Definition~1.3.2]{baer+ginoux}). Then for any~$x, y \in \Omega$ there is a
unique (unparametrized) geodesic~$\gamma$ in~$\Omega$ joining~$y$ and~$x$. We denote the squared length of this geodesic by
\[ \Gamma(x,y) = g \big( \exp_y^{-1}(x), \exp_y^{-1}(x) \big) \]
(note that~$\Gamma$ is positive for timelike and negative for spacelike separation).

\subsection{Ansatz for the Regularized Hadamard Expansion}
For the calculations it is most convenient to set~$X_{-1} = U_0$, $X_n = V_n$
and~$Y_n = W_n + U_{n-1}$. Then the Hadamard expansion~\eqref{Tform}--\eqref{UVWexpand} takes the form
\[ 
T(x,y)  = \lim_{\varepsilon \searrow 0} \bigg( \frac{ X_{-1}}{   \Gamma_\varepsilon} 
+  \log \Gamma_\varepsilon \sum_{n=0}^{\infty} X_n \:\Gamma_\varepsilon^n
+ \sum_{n=0}^{\infty} Y_n \:\Gamma_\varepsilon^n \bigg) \:. \]
For the regularized Hadamard expansion~\eqref{reghadamard} we
insert functions~$i \varepsilon \f_{[n]}$ and~$i \varepsilon \f_{\{n\}}$ into the factors~$\Gamma$
(see~\eqref{Gammaf}). Moreover, we choose the functions~$X_n$ and~$Y_n$
according to~\eqref{XA} and~\eqref{YB}.
The advantage of~\eqref{reghadamard} is that it contains information on the dynamics of the
regularization.
In view of later generalizations, it is preferable to replace the square of the mass
by an arbitrary smooth function~$\mu(x)$. Thus we must solve the equation
\beq \label{Teq}
\big( \Box_x +\mu(x) \big)\,T^\varepsilon(x,y) = 0 \:,
\eeq
order by order in~$n$ and modulo error terms of the form~\eqref{fehler}.

\subsection{The Transport Equations} \label{sectransport}
$\;\;\;$ Evaluating the regularized Hadamard expan\-sion~\eqref{reghadamard}
in the Klein-Gordon equation~\eqref{Teq} gives rise to
families of transport equations. In order not to distract from the main
constructions, we here state these transport equations and refer to
the detailed computations to Appendix~\ref{appendixA}.
For the functions~$X_\bullet$ and~$Y_\bullet$, which already appear in the
unregularized Hadamard expansion, we obtain the usual transport equations:
\begin{Prp} \label{prphadamard1}
The functions~$A_\bullet$ and~$B_\bullet$ must satisfy 
for all~$n \in \N_0$ the transport equations
\begin{align}
2 \,\langle \nabla \Gamma, \nabla A_{-1} \rangle &= -\big(\Box\, \Gamma -8 \big)\:A_{-1} \label{Xm1} \\
2 \,\langle \nabla \Gamma, \nabla A_0 \rangle &= -\big(\Box\, \Gamma -4\big)\:A_0 - (\Box + \mu) A_{-1} \label{X0} \\
2\,\langle \nabla \Gamma, \nabla A_{n+1} \rangle &= -\big(
\Box\, \Gamma + 4n \big) \:A_{n+1}- \frac{(\Box + \mu) A_n}{n+1} \label{Xnp1} \\
2 \,\langle \nabla \Gamma, \nabla B_{n+1} \rangle 
&=-\big(\Box\, \Gamma + 4n \big)\:B_{n+1} - \frac{(\Box + \mu) B_n}{n+1} - 4 \, A_{n+1} + \frac{(\Box + \mu) A_n}{(n+1)^2} \label{Ynp1} \:,
\end{align}
where~$B_0$ is an undetermined smooth function.
\end{Prp} \noindent

The point of interest is that we also get transport equations for the functions
describing the regularization:
\begin{Prp} \label{prptransport2}
The functions~$\f_\bullet$, $a_\bullet$ and~$b_\bullet$
must satisfy  for all~$n \in \N_0$ the transport equations
\begin{align}
\f &:= \f_{[-1]} = \f_\bullet \qquad \text{and}
\qquad 
2\, \langle \nabla \Gamma, \nabla \f \rangle = 4\,\f \label{fm1} \\
2 \,\langle &\nabla \Gamma, \nabla \fa_{-1} \rangle - \big(8 - \Box\, \Gamma \big) \:\fa_{-1} \notag \\
&= - 2 \, \langle \nabla \f, \nabla A_{-1} \rangle  - (\Box\, \f)\:A_{-1} \label{Am1} \\
2 \,\langle &\nabla \Gamma, \nabla \fa_{0} \rangle - \big(4 - \Box\, \Gamma \big) \:\fa_{0} \notag \\
&= - 2 \,\langle \nabla \f, \nabla A_0 \rangle - (\Box\, \f) \:A_0
- \big(\Box\, \fa_{-1} + \mu(x)\, \fa_{-1}\big) \label{A0} \\
2\,\langle &\nabla \Gamma, \nabla \fa_{n+1} \rangle + \big( 4n + \Box\, \Gamma \big)\: \fa_{n+1} \notag \\
&= -2 \,\langle \nabla \f, \nabla A_{n+1} \rangle - (\Box\, \f)\:A_{n+1}
 - \frac{1}{n+1} \:\big(\Box\, \fa_{n} + \mu(x)\, \fa_{n} \big) \label{An} \\
2 \,\langle &\nabla \Gamma, \nabla \fb_{n+1} \rangle
+ \big(4n + \Box\, \Gamma \big) \:\fb_{n+1} \notag \\
&= -2 \,\langle \nabla \f, \nabla B_{n+1} \rangle - (\Box\, \f)\: B_{n+1} 
- \frac{1}{n+1}\: \big(\Box\, \fb_{n} + \mu(x)\, \fb_{n} \big) \label{Bn1} \\
&\quad\, + \frac{1}{(n+1)^2}\: \Big( \Box\, \fa_{n} + \mu(x)\, \fa_{n} \Big) - 4\: \fa_{n+1} \:, \label{Bn2}
\end{align}
where~$\fb_{0}$ is an undetermined function.
\end{Prp} \noindent
Note that the right side of these equations are determined by the solutions of the previous transport equations
and can therefore be treated as given inhomogeneities.

At this stage, we can make the error terms of our expansion more precise.
In view of~\eqref{fm1}, all the functions~$\Gamma_\bullet$ coincide. Therefore, we can simplify the
notation by setting
\[ \Gamma_\varepsilon := \Gamma_{[-1]}\:. \]
Moreover, we can improve the error term~\eqref{fehler} to
\beq \label{fehlerneu}
\bigg(1 + \O\Big( \frac{\varepsilon^2}{\Gamma_\varepsilon} \Big) \bigg) \:.
\eeq
This is the error term which we will work with in what follows.

\subsection{Solving the Transport Equations} \label{secsolve}
We now explain how to solve transport equations.
Our task is to show existence of solutions, and to specify in which sense the
regularized Hadamard expansion is unique.
We begin with the first group of transport equations~\eqref{Xm1}--\eqref{Ynp1}.
These equations do not involve the regularization, and they
appear in exactly the same way in the standard Hadamard expansion (see for example~\cite[Section~2]{baer+ginoux}).
In order to analyze existence and uniqueness, we evaluate these equations for~$x=y$.
This gives for~$n \in \N_0$ the conditions
\begin{align*}
\eqref{Xm1}: \qquad 0 &= -\big(8-8 \big)\:A_{-1} \\
\eqref{X0}: \qquad 0 &= -\big(8-4 \big)\:A_0 - (\Box + \mu) A_{-1} \\
\eqref{Xnp1}: \qquad 0 &= - \big( 8+4n \big)\:A_{n+1} - \frac{(\Box + \mu) A_n}{n+1} \\
\eqref{Ynp1}: \qquad 0 &=-\big(8+4n \big)\:B_{n+1} - \frac{(\Box + \mu) B_n}{n+1} - 4 \, A_{n+1} + \frac{(\Box + \mu) A_n}{(n+1)^2} \:.
\end{align*}
The condition resulting from~\ref{Xm1} is trivially satisfied. Obviously, multiplying a solution~$A_{-1}$
of the transport equation~\eqref{Xm1} by any complex number again gives a solution of the transport equation.
This corresponds to the freedom of multiplying the distribution~$T^\varepsilon(x,y)$ by a prefactor.
In order to fix this freedom, we choose~$A_{-1}$ as the square root of the
van Vleck-Morette determinant (see for example~\cite{moretti} or~\cite[eqns~(1.9) and~(2.4)]{baer+ginoux}),
\[ A_{-1} = \vleck \:, \]
which in normal coordinates around $y$ is given by 
\[ \vleck(x,y)=|\det(g(x))|^{-\frac{1}{4}} \:. \]
Being smooth, the function~$B_0$ is undetermined.
The conditions for the other transport equations~\eqref{X0}--\eqref{Ynp1}
determine the initial conditions~$A_0(x,x)$, $A_{n+1}(x,x)$ as well as~$B_{n+1}(x,x)$
(for~$n=0,1,2\ldots$). As a consequence, the resulting transport equations have unique
solutions. For more details and explicit formulas we refer to~\cite[Sections~2.2 and~2.3]{baer+ginoux}.

We now turn attention to the transport equations~\eqref{fm1}--\eqref{Bn2} for the functions~$\f$,
$a_\bullet$ and~$b_\bullet$.
The transport equation~\eqref{fm1} has a structure which is very similar
to that of the usual Hadamard transport equations in Proposition~\ref{prphadamard1}.
However, there is a major difference: The requirement~$\lim_{\varepsilon \searrow 0} T^\varepsilon =T$
implies that the function~$\f(x,y)$ must have the same sign as $\frak{t}(x)-\frak{t}(y)$.
In particular, this leads to the condition
\beq \label{fzero}
\f(x,x) = 0 \:.
\eeq
Therefore, the question is whether the transport equations in Proposition~\ref{prptransport2}
admit solutions with boundary conditions~\eqref{fzero}, and to which extent they are unique.
We begin with the uniqueness problem:
\begin{Lemma} \label{lemmaunique}
The solutions of the transport equations of Proposition~\ref{prptransport2}
subject to the initial condition~\eqref{fzero} are unique up to contributions of the following form:
\begin{align}
&\text{$\f$ is unique up to a multiplicative constant} \label{fm1u} \\
&\text{$\fa_{-1}$ is unique up to a multiple of~$A_{-1}$} \label{Am1u} \\
&\text{$\fb_{0}$ is arbitrary} \label{B0u} \\
&\text{$\fa_{0}, \fa_{1}, \ldots$ are unique} \label{Au} \\
&\text{$\fb_{0}, \fb_{1}, \ldots$ are unique} \label{Bu} \:.
\end{align}
\end{Lemma}
\Proof The statement~\eqref{fm1u} follows immediately from the fact that~\eqref{fm1} is
a homogeneous ODE of first order. In order to prove~\eqref{Am1u}, we note that
the homogeneous part of~\eqref{Am1} coincides with the transport equation for~$A_{-1}$
in~\eqref{Xm1}, whose general solution is a multiple times the van Vleck-Morette determinant.
The statement~\eqref{B0u} is obvious because
there is no transport equation for~$\fb_{0}$.
The uniqueness problem for the resulting transport equations~\eqref{A0}--\eqref{Bn2}
can be solved exactly as for the transport equations~\eqref{X0}--\eqref{Ynp1} by noting
that the factor~$\Box\, \Gamma + 4n$ is non-zero on the diagonal for~$n \geq -1$.
This concludes the proof.
\QED

Before delving into the existence problem, we point out that it suffices to solve the
transport equations along all {\em{null geodesics}}. Indeed, doing so determines
the functions~$\f$, $a_{-1}$, $a_n$ and~$b_n$ for points~$x$ and~$y$ with lightlike separation.
Extending these functions smoothly to~$\scrM \times \scrM$, the resulting freedom gives rise to
errors which are taken care of by the transport equations in the subsequent order of the expansion.

For a null geodesic~$\gamma$, the transport equation~\eqref{fm1} means that~$\f$
is an {\em{affine parameter}} along~$\gamma$.
This observation, which was first made in~\cite[Appendix~B]{dgc},
is verified as follows:
Let~$\Gamma$ be a null geodesic with~$\gamma(0)=y$ and~$\gamma(\tau)=x$.
Then
\[ \nabla_x \Gamma(x,y) = 2 \tau\: \frac{d}{d\tau} \gamma(\tau) \]
(note that the right side does not depend on the multiplicative parametrization of~$\gamma$).
Therefore, we may write the transport equation~\eqref{fm1} as
\[  \tau \:\dot{\gamma}^j(\tau)  \:\partial_j \f \big( \gamma(\tau) \big) = \f \big( \gamma(\tau) \big) \:, \]
implying that
\[ \tau\: \frac{d}{d\tau} \f \big( \gamma(\tau) \big) = \f \big( \gamma(\tau) \big) \:. \]
Setting~$h(\tau) =  h(\gamma(\tau))$, we obtain the ordinary differential equation
\[ \tau\: \frac{d}{d\tau} h(\tau) = h(\tau) \:, \]
having the general solution~$h(\tau) = c \tau$ with a free real parameter~$c$.
We conclude that~$\f$ has the general form
\[ \f \big( \gamma(\tau) \big) = c \tau \:. \]
Hence along a null geodesic through~$y$, the function~$\f(.,y)$ simply is an affine parameter
along this geodesic with~$\f(y,y)=0$.
The non-uniqueness~\eqref{fm1u} corresponds precisely to the freedom in choosing a
distinguished parametrization of each null geodesic.

We now give a systematic procedure for solving the transport equations of Proposition~\ref{prptransport2}.
To this end, given~$y \in \scrM$ we choose a null geodesic~$\gamma$ through~$y$
and consider a point~$x$ along this geodesic.
As already mentioned above, $\f$ is an affine parameter along~$\gamma$.
Using~\eqref{fzero}, it is unique up to a multiplicative constant.
We always choose the parametrization of the geodesics
to depend smoothly on the null geodesic.
Then the resulting function~$\f(x,y)$ is smooth in~$x$ and~$y$ (where~$x$ is always
on the light cone centered at~$y$), provided that~$x$ stays away from~$y$.
On the diagonal~$x=y$, however, the function~$\f$ is only continuous, but it will in general {\em{not}} be
smooth. We also point out that this procedure determines~$\f$ only on the light cone
(i.e.\ for pairs of points~$x,y$ with~$\Gamma(x,y)=0$). Away from the light cone,
we are free to modify~$\f$ arbitrarily.

Solving the ODE~\eqref{Am1} is more subtle. 
Using that~$\nabla \Gamma(x,x)=0$
and that~$\f$ vanishes according to~\eqref{fzero}, 
evaluating this transport equation at~$x=y$ gives rise to the algebraic conditions
\beq \label{fA1}
- 2 \, \langle \nabla \f, \nabla A_{-1} \rangle  - (\Box\, \f)\:A_{-1} = 0 \:,
\eeq
which must hold at~$x=y$.
At this point, we can make use of the fact that the function~$\f$ need not be smooth
on the diagonal, and that it can be changed arbitrarily away from the light cone.
Using this freedom, one can indeed arrange that the function vanishes
on the light cone in a neighborhood of~$y$, as we now explain: We want to arrange that~\eqref{fA1}
holds on the light cone centered at~$y$. In order to eliminate the first order term,
it is preferable to write this equation equivalently as
\beq \label{PDE}
\Box \big( \f \,A_{-1} \big) = \big(\f\,A_{-1} \big)\: \frac{\Box A_{-1}}{A_{-1}} \:.
\eeq
This partial differential equation is analyzed most conveniently in
light cone coordinates $(u,v,\vartheta, \varphi)$ around~$y$, where~$u$ vanishes on the
future light cone and~$v$ vanishes on the past light cone (both centered ab~$y$).
We choose double null coordinates at~$y$. To this end, we first choose a Gaussian
coordinate system~$(t,x,y,z)$ where~$g_{ij}(y)= \text{diag}(1,-1,-1,-1)$ is the Minkowski metric.
Choosing standard coordinates~$(r, \vartheta, \varphi)$ and null coordinates~$u=(t-r)/2$ and~$v=(t+r)/2$,
the line element becomes
\[ ds^2 = \Big( 4\,du\, dv + (v-u)^2\, g_{S^2}(\vartheta, \varphi) \Big) \big( 1+ \O(u) + \O(v) \big) \:, \]
so that the Laplacian at~$y$ takes the form
\[ \Box = \frac{\partial}{\partial u} \frac{\partial}{\partial v}+\frac{1}{v-u}\:\bigg(
\frac{\partial}{\partial v} -\frac{\partial}{\partial u} \bigg) - \frac{1}{(v-u)^2}\: \Box_{S^2}\:. \]
The function~$\f\, A_{-1}$ is already determined on the light cone.
It grows linearly in~$v-u$ and is smooth in the angular variables. Hence the angular part of the
Laplacian has at most a simple pole at the origin, i.e.\ on the future light cone
\[ \frac{1}{(v-u)^2}\: \Box_{S^2} \big( \f \,A_{-1} \big) \big|_{u=0} \lesssim \frac{1}{v} \:. \]
Therefore, choosing the $u$-derivative on the upper light cone such that
\[ \partial_u \big( \f\, A_{-1} \big)\big|_{u=0} \lesssim \frac{1}{v} \:, \]
one can arrange that the equation~\eqref{PDE} is satisfied on the future light cone in a neighborhood of~$y$.
Similarly, by choosing the~$v$-derivative on the lower light cone appropriately, we can arrange
that~\eqref{PDE} holds on the past light cone in a neighborhood of~$y$.
Finally, we extend the function~$\f \,A_{-1}$ respecting its prescribed first derivatives on the light cone
such that it is smooth for~$x \neq y$ and continuous at~$x=y$.

In this way, we have arranged that~\eqref{fA1} holds on the light cone
in a neighborhood of~$y$.
As a consequence, the transport equation~\eqref{Am1}
clearly admits a solution which vanishes at~$x=y$.

The remaining transport equations~\eqref{A0}--\eqref{Bn2} can be
solved exactly as the usual transport equations~\eqref{X0}--\eqref{Ynp1}.

In order to complete the proof of Theorem~\ref{thmmain}, it remains to 
analyze the symmetry properties of the solution~$\f$.
\begin{Lemma} The function~$\f$ is antisymmetric~\eqref{fantisymm}.
\end{Lemma}
\Proof We again fix~$y$ and choose a null geodesic~$\gamma$ through~$y$.
Then, as explained above, the function~$\f(x,y)$ is an affine parameter~$\tau$
of the geodesic, i.e. $\f(x,y)= \tau(x) - \tau(y)$. This shows that~$\f(x,y) = -\f(y,x)$,
concluding the proof.
\QED

\section{Propagation of Regularized Singularities} \label{secprop}
So far, the regularized Hadamard expansion was performed locally
in a convex neighborhood~$\Omega \subset \scrM$. In this section we
specify assumptions under which the expansion is globally well-defined.
Moreover, we show that the regularized Hadamard form is preserved by the time evolution.
By a {\em{global expansion}} we mean an expansion which can be performed
in every convex neighborhood and is uniquely determined, up to smooth
contributions and well-defined error terms.
In particular, the expansions in two convex neighborhoods should coincide
in the intersection of these neighborhoods.

The first step in the construction is to determine the function~$\f$ globally.
To this end, we distinguish parametrizations of null geodesic as follows:
Let~$\gamma(\tau)$  be a parametrized future-directed
null geodesic defined on an open interval~$I \subset \R$.
Such a geodesic can be reparametrized in several ways. One obvious freedom is to
change the parameter by an {\em{additive constant}},
\beq \label{addchange}
\tilde{\gamma}(\tau) := \gamma(\tau+c) \qquad \text{with} \qquad
\tau \in \tilde{I} := I - c \:.
\eeq
Moreover, one may reparametrize by a non-negative {\em{multiplicative constant}}~$\lambda$, i.e.
\beq \label{multchange}
\tilde{\gamma}(\tau) := \gamma(\lambda \tau) \qquad \text{with} \qquad
\text{$\tau \in \tilde{I} := I/\lambda$ and~$\lambda>0$} \:.
\eeq
This freedom scales the velocity vector by~$\dot{\tilde{\gamma}}(\tau) = \lambda\, \dot{\gamma}(\lambda \tau)$.
The additive and multiplicative reparametrizations indeed exhaust the freedom to
change parametrizations of null geodesics.

In the next definition we distinguish the multiplicative parametrization of every null geodesic.
Clearly, it suffices to consider maximal geodesics (i.e.\ geodesics which are inextendible),
because all other null geodesics can be obtained by restriction
to a smaller parameter domain.
\begin{Def} \label{defL}
We introduce a set of parametrized future-directed maximal null geo\-desics
\[ 
\scrL = \big\{ (\gamma, I) \:|\: \text{$\gamma \::\: I \rightarrow \scrM$ is a
future-directed maximal null geodesic} \big\} \]
with the following properties:
\begin{itemize}[leftmargin=2em]
\item[\rm{(a)}] For every~$(\gamma, I) \in \scrL$, reparametrizing by an additive constant~\eqref{addchange}
again gives a geodesic in~$\scrL$.
\item[\rm{(b)}] For every maximal null geodesic~$\gamma(\tau)$ in~$\scrM$, there is exactly one~$\lambda>0$
such that the multiplicative reparametrization~\eqref{multchange} gives a
geodesic~$(\tilde{\gamma}, \tilde{I}) \in \scrL$.
\end{itemize}
\end{Def} \noindent
For any space-time point~$p \in \scrM$ we introduce the set
\[ 
D_p\scrL = \big\{ \dot{\gamma}(\tau) \:\big|\: \text{$(\gamma, I) \in \scrL$, $\tau \in I$
and $\gamma(\tau) = p$} \big\} \subset T_p\scrM \:. \]
Clearly, this set is a subset of the light cone centered at~$p$, i.e.
\[ D_p\scrL \subset \{ u \in T_p\scrM \,|\, \la u,u \ra_p = 0 \} \:. \]

Given~$\scrL$, the function~$\f$ can be chosen canonically as follows.
Let~$\Omega$ be a convex neighborhood and let~$x,y \in \Omega$ be two space-time points
with lightlike separation. Then there is a unique
unparametrized null geodesic joining~$x$ and~$y$. According to Definition~\ref{defL}~(ii)
there is a corresponding parametrized null geodesic~$\tilde{\gamma}(\tilde{\tau})$ in~$\scrL$, which is unique up
to additive reparametrizations. We choose~$\tau_x$ and~$\tau_y$ with~$\tilde{\gamma}(\tau_x)=x$
and~$\tilde{\gamma}(\tau_y)=y$ and set
\beq \label{ftaudef}
\f(x,y) = \tau_x - \tau_y \:.
\eeq
We refer to this choice as the function~$\f$ {\em{induced by~$\scrL$}}.

The next question is whether a family~$\scrL$ exists which has the desired smoothness
properties. An easy method for constructing~$\scrL$ is to choose the parametrizations on a
Cauchy surface:
\begin{Lemma} Let~$\scrN$ be a Cauchy surface. For any~$p \in \scrN$ we let~$\lambda$
be a positive smooth function on the sphere bundle,
\[ \lambda \::\: S\scrN \subset T\scrN \rightarrow \R^+ \]
(thus~$\lambda$ maps any vector in~$T_x\scrN$ which has length one with respect to the
induced Riemannian metric to a positive number). Then there is a unique family
of parametrized null geodesics~$\scrL$ as in Definition~\ref{defL} such that for
every parametrized null geodesic in~$\scrL$ with~$\gamma(0) \in \scrN$,
\beq \label{nullcond}
\dot{\gamma}(0) \in \big\{ \lambda(n)\: \big(\nu + n \big) \;\big|\; n \in S_{\gamma(0)}\scrN \big\} \:,
\eeq
where~$\nu$ denotes the unit future normal on~$\scrN$ in~$\gamma(0)$
(and we consider~$n$ as a vector in~$S_{\gamma(0)}\scrN \subset T_{\gamma(0)}\scrM$).
\end{Lemma}
\Proof 
Since every maximal null geodesic intersects the Cauchy surface~$\scrN$ exactly once,
the condition~\eqref{nullcond} determines the multiplicative parametrization uniquely.
\QED
We point out that this construction applies even if there are conjugate points,
implying that two points~$x,y \in \scrM$ may be joined by two different geodesics.
In this case, the family~$\scrL$ can still be constructed, because the geodesic equation
is globally well-defined. The function~$\f$, on the other hand, is only defined locally in convex neighborhoods,
but since it is induced by~$\scrL$ via~\eqref{ftaudef}, it is uniquely determined by~$\scrL$
in any convex neighborhood.

The basic difficulty is to control the error of the expansion in the time evolution.
It turns out that error terms of the form~\eqref{fehler} are not suitable.
Instead, we must work with the expansion parameter~$\varepsilon^2/\Gamma_\varepsilon$.
Moreover, it does not suffice to consider the terms of zero order in this expansion,
but instead we must consider an expansion in powers of~$\varepsilon^2/\Gamma_\varepsilon$
up to a fixed order~$k_{\max}$ (to be specified below). This leads us to the ansatz
\beq \begin{split}
R^\varepsilon(x,y) &= \sum_{k=0}^{k_{\max}} 
\varepsilon^{2k} \:\Gamma_\varepsilon^{-1-k} \:\big( A_{-1,k} + i \varepsilon \,\fa_{-1,k} \big) \\
&\quad\, + \sum_{n=0}^\infty \sum_{k=0}^{k_{\max}} \varepsilon^{2k} \:\Gamma_\varepsilon^{n-k}
\Big[ \big(A_{n,k} + i \varepsilon \,\fa_{n,k} \big) \: \log \Gamma_\varepsilon 
+ \big( B_{n,k} + i \varepsilon \,\fb_{n,k} \big) \Big] \:,
\end{split} \label{RepsnoN}
\eeq
where all coefficient functions are continuous as well as smooth away from the diagonal.
In order to bypass convergence issues, we always truncate this series by setting
\beq \begin{split}
R^\varepsilon_N(x,y) &= \sum_{k=0}^{k_{\max}} 
\varepsilon^{2k} \:\Gamma_\varepsilon^{-1-k} \:\big( A_{-1,k} + i \varepsilon \,\fa_{-1,k} \big) \\
&\quad\, + \sum_{n=0}^N \sum_{k=0}^{k_{\max}} \varepsilon^{2k} \:\Gamma_\varepsilon^{n-k}
\Big[ \big(A_{n,k} + i \varepsilon \,\fa_{n,k} \big) \: \log \Gamma_\varepsilon 
+ \big( B_{n,k} + i \varepsilon \,\fb_{n,k} \big) \Big] \:,
\end{split} \label{RepsN}
\eeq
where~$N \in \N$ is a parameter which we will be allowed to choose arbitrarily large.
Since the factor~$\log \Gamma_\varepsilon$ is never multiplied by negative powers of~$\Gamma_\varepsilon$,
we demand that
\[ A_{n,k} = \fa_{n,k} = 0 \qquad \text{if~$k>n$}\:. \]
Moreover, we fix the freedom to add smooth solutions by setting the constant terms to zero,
\[ B_{n,k} = \fb_{n,k} = 0 \qquad \text{if~$k=n$}\:. \]

Given a compact subset~$K \subset \Omega$, we choose a continuous distance function~$d$ on~$K$
and introduce the compact set
\[ {\mathscr{P}}_{K, d} := \big\{(x,y) \in K \times K \:\big|\: d(x,y) \geq 1 \big\} \:. \]
\begin{Def} The series~$R^\varepsilon(x,y)$ approximates the
bi-solution~$T^\varepsilon(x,y)$ {\bf{to the order}}~$\O^s(\varepsilon^2)$ if
for any compact subset~$K \subset \Omega$, for any
continuous distance function~$d$ on~$K$ and for all sufficiently large~$N \in \N$,
\beq \label{genex}
\big( T^\varepsilon - R^\varepsilon_N \big) \big|_{{\mathscr{P}}_{K, d}}
= C^s({\mathscr{P}}_{K, d})\: \O\big(\varepsilon^2 \big)
+ C^N({\mathscr{P}}_{K, d}) \:.
\eeq
\end{Def}

The order of differentiability~$s$ of the error term is determined by the parameter~$k_{\max}$.
The precise connection is obtained as follows.
If~$T^\varepsilon$ had an exact expansion of the form~\eqref{RepsN} with~$N=k=\infty$,
then the leading error terms of the difference~$T^\varepsilon - R^\varepsilon_N$ would be of the form
\[ \sim \Gamma_\varepsilon^{N+1}\, \big(1+\log \Gamma_\varepsilon \big)
\qquad \text{and} \qquad \sim \varepsilon^{2 k_{\max}+2} \:\Gamma_\varepsilon^{-2-k_{\max}} \:. \]
Differentiating in~$x$ decreases the order of~$\Gamma_\varepsilon$ at most by one. Therefore,
for any multi-index~$\alpha$ with~$|\alpha|=p$,
\[ \Big| \frac{\partial^p}{\partial x^\alpha} \:\Gamma_\varepsilon^{N+1}\, \big(1+\log \Gamma_\varepsilon \big) \Big| 
\lesssim |\Gamma_\varepsilon|^{N+1-p}\: \big(1 + |\log \Gamma_\varepsilon| \big)
\lesssim |\Gamma_\varepsilon|^{N-p} \:, \]
showing that the first error term is of the class
\[ \Gamma_\varepsilon^{N+1}\, \big(1+\log \Gamma_\varepsilon \big) \in C^N({\mathscr{P}}_{K, d}) + \O(\varepsilon^2)\: C^{N-1}({\mathscr{P}}_{K, d})\:. \]
Differentiating the second error term, we obtain similarly
\[ \Big| \frac{\partial^p}{\partial x^\alpha} \:\varepsilon^{2 k_{\max}+2} \:\Gamma_\varepsilon^{-2-k_{\max}} \Big| 
\lesssim \frac{\varepsilon^{2 k_{\max}+2}}{\Gamma_\varepsilon^{2+k_{\max}+p}} 
\lesssim \frac{\varepsilon^{2 k_{\max}+2}}{(\varepsilon\, |\f(x,y)|)^{2+k_{\max}+p}} 
= \frac{\varepsilon^{k_{\max}- p}}{|\f(x,y)|^{2+k_{\max}+p}} \:. \]
Hence the second error term is of the class
\[ \varepsilon^{2 k_{\max}+2} \:\Gamma_\varepsilon^{-2-k_{\max}}  \in \O(\varepsilon^2)\:
C^{k_{\max}-2}({\mathscr{P}}_{K, d})\:. \]
Comparing with~\eqref{genex}, we are led to choosing
\beq \label{kchoice}
k_{\max} = s+2 \qquad \text{and} \qquad N \geq s+1 \:.
\eeq

\begin{Def} \label{defregHad} The bi-solution~$T^\varepsilon$ is of {\bf{regularized Hadamard form
on~$U \subset \scrM$}} to the order~$\O^s(\varepsilon^2)$
if every~$x_0 \in U$ has a convex neighborhood~$\Omega \subset \scrM$
such that there is a series~$R^\varepsilon$ of the form~\eqref{RepsnoN} defined on~$\Omega \times \Omega$
which approximates~$T^\varepsilon(x,y)$ to the order~$\O^s(\varepsilon^2)$.
\end{Def}

\begin{Thm} \label{thmpropagate}
Assume that the bi-solution~$T^\varepsilon(x,y)$ of the Klein-Gordon equation
is of regularized Hadamard form
on the Cauchy surface~$\scrN$ to the order~$\O^s(\varepsilon^2)$ with
\beq \label{sbound}
s \geq 2 + \Big[ \frac{m}{2} \Big]
\eeq
(where~$m = \dim \scrN$ is the spatial dimension).
Then it is of regularized Hadamard form on all of~$\scrM$ to the order~$\O^{r}(\varepsilon^2)$, where
\[ 
r = s - 2 - \Big[ \frac{m}{2} \Big] \:. \]
\end{Thm}
We now enter the proof of this theorem, which will be completed at the end of this section.
We begin with a lemma on the structure of the regularized Hadamard expansion~\eqref{RepsN}
and the corresponding transport equations:
\begin{Lemma} \label{lemmastructure} Evaluating the equation
\[ \big(\Box_x + \mu(x) \big) R^\varepsilon_N(x,y) = 0 \]
in increasing order in~$n$ and~$k$,
one obtains transport equations for the unknown functions~$A_{n,k}$, $B_{n,k}$
and~$\fa_{n,k}$, $\fb_{n,k}$ of the general form
\beq \label{genform}
\langle \nabla \Gamma, \nabla g \rangle = \cdots \:,
\eeq
where the right side depends on~$g$ as well as on previously computed functions.
The error term is of the class
\beq \label{sval}
\big(\Box_x + \mu(x) \big) R^\varepsilon_N(x,y) \:\in\: C^{s-1}({\mathscr{P}}_{K, d})\, \O\big( \varepsilon^2 \big) + C^N({\mathscr{P}}_{K, d})
\eeq
(where we again chose~$k_{\max}$ according to~\eqref{kchoice}).
\end{Lemma}
\Proof The transport equations for~$k=0$ are derived in Appendix~\ref{appendixA}.
The error term involves a power of~$\varepsilon^2/\Gamma_\varepsilon$
(see~\eqref{transans} and the explicit formula for~$R(x,y)$ at the end of the appendix).
Therefore, we can proceed inductively in~$k$, giving rise to transport equations
of the form~\eqref{genform}.
It remains to determine the error terms. We begin with the error terms due to the
truncation in~$n$. For~$k=0$, we obtain an error term of the form
\[ \big(\Box_x + \mu(x) \big) R^\varepsilon_N(x,y) \sim \Gamma_\varepsilon^N\: \log \Gamma_\varepsilon \:. \]
In order to determine the regularity of the right side, we again use that differentiating
decreases the order of~$\Gamma_\varepsilon$ at most by one. Therefore,
for any multi-index~$\alpha$ with~$|\alpha|=p$,
\[ \Big| \frac{\partial^p}{\partial x^\alpha} \big(\Box_x + \mu(x) \big) R^\varepsilon_N(x,y) \Big| 
\lesssim |\Gamma_\varepsilon|^{N-p}\: \big(1 + |\log \Gamma_\varepsilon| \big) \\
\lesssim |\Gamma_\varepsilon|^{N-p-1} \:, \]
showing that the error term is of the class~$C^{N-1}$. Since increasing~$k$ gives
a scaling factor~$\varepsilon^2/\Gamma_\varepsilon$, we conclude that the error terms
of the truncation of~$n$ have the regularity
\[ \big(\Box_x + \mu(x) \big) R^\varepsilon_N(x,y) \:\in\: C^N + \varepsilon^2\: C^{N-1}
+ \cdots + \varepsilon^{2 k_{\max}}\: C^{N-k_{\max}} \:. \]

In the case~$k_{\max}=0$, the leading error term coming from the truncation in~$N$
is computed in Appendix~\ref{appendixA} to be
\[ \big(\Box_x + \mu(x) \big) R^\varepsilon_N(x,y) \sim \frac{\varepsilon^2}{\Gamma_\varepsilon^3} \]
(see again~\eqref{transans} and the explicit formula for~$R(x,y)$ at the end of the appendix).
Since incrementing~$k$ gives rise to a a factor~$\varepsilon^2/\Gamma_\varepsilon$,
the leading error for general~$k_{\max}$ is given by
\[ \big(\Box_x + \mu(x) \big) R^\varepsilon_N(x,y) \sim \frac{\varepsilon^{2+2 k_{\max}}}{\Gamma_\varepsilon^{3+k_{\max}}} \:. \]
Therefore, for any any multi-index~$\alpha$ with~$|\alpha|=p$,
\[ \Big| \frac{\partial^p}{\partial x^\alpha} \big(\Box_x + \mu(x) \big) R^\varepsilon_N(x,y) \Big| 
\lesssim \frac{\varepsilon^{2+2 k_{\max}}}{\Gamma_\varepsilon^{3+k_{\max}+p}} 
\lesssim \frac{\varepsilon^{2+2 k_{\max}}}{(\varepsilon\, |\f(x,y)|)^{3+k_{\max}+p}} 
= \frac{\varepsilon^{k_{\max}- p - 1}}{|\f(x,y)|^{3+k_{\max}+p}} \:. \]
We conclude that, away from diagonal,
\[ \big(\Box_x + \mu(x) \big) R^\varepsilon_N(x,y) \:\in\:
C^{s-1}\, \O\big( \varepsilon^2 \big) \]
with~$s$ according to~\eqref{kchoice}.
Collecting all the error terms gives the result.
\QED

\begin{Lemma} \label{lemmacauchy}
Let~$(\scrN_t)_{t \in \R}$ be a foliation of~$\scrM$ by Cauchy surfaces.
Given~$s$ satisfying~\eqref{sbound}, the Cauchy problem
\[ (\Box_x + \mu(x)) \phi(x) = g(x) \in W^{s-1,\,2}_\text{\rm{loc}}(\scrN_0) \]
with initial values
\[ \phi|_\scrN = \phi_0 \in W^{s,\,2}_\text{\rm{loc}}(\scrN_0)\:, \qquad \partial_\nu \phi|_\scrN = \phi_1 \in W^{s-1,\,2}_\text{\rm{loc}}(\scrN_0) \]
has a unique global weak solution. Its restriction to any other Cauchy surface~$\scrN_t$ is of
the class~$W^{s,2}_\text{\rm{loc}}(\scrN_t)$.
\end{Lemma}
\Proof Due to finite propagation speed, it clearly suffices to solve the Cauchy problem in
a local chart. Rewriting the equations as a symmetric hyperbolic system for~$\psi = (\partial_t \phi, \nabla \phi, \phi)$,
the initial data and the inhomogeneity are in the Sobolev space~$W^{s-1,\,2}_\text{loc}(\R^m)$.
Therefore, there is a unique global solution in this Sobolev space
(see for example~\cite[Chapter~5.3]{john} or~\cite[Chapter~9]{intro}).
Moreover, since the first derivatives of the function~$\phi$ are in~$W^{s-1,\,2}_\text{loc}(\R^m)$,
this function is even in~$W^{s,2}_\text{loc}(\R^m)$.
\QED

\Proof[Proof of Theorem~\ref{thmpropagate}]
Applying the Sobolev embedding~$W^{s,2}_{\text{loc}}(\R^m) \hookrightarrow C^r(\R^m)$
(see for example~\cite[Section~5.6.3]{evans}) it suffices to estimate $T^\varepsilon - R^\varepsilon_N$
in the Sobolev space~$W^{s,2}$ on Cauchy surfaces.

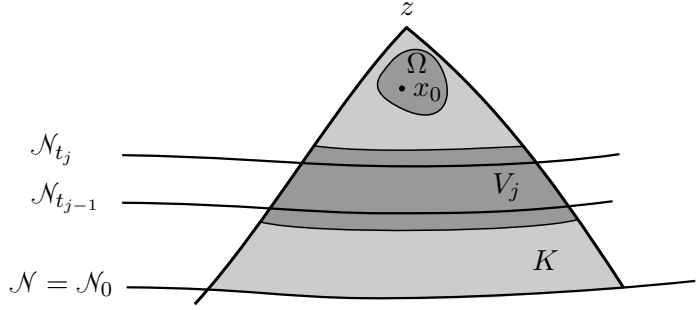
\begin{figure}
%
\psscalebox{1.0 1.0} 
{
\begin{pspicture}(0.8,-2.0067608)(9.116561,2.0067608)
\definecolor{colour0}{rgb}{0.8,0.8,0.8}
\definecolor{colour1}{rgb}{0.6,0.6,0.6}
\pspolygon[linecolor=colour0, linewidth=0.02, fillstyle=solid,fillcolor=colour0](5.29,1.6717609)(5.525,1.4667609)(6.035,0.9667609)(6.365,0.6267609)(6.685,0.2517609)(7.19,-0.3682391)(7.5,-0.8032391)(7.785,-1.1932391)(7.995,-1.5182391)(8.15,-1.7832391)(7.435,-1.8132391)(6.46,-1.8932391)(5.465,-1.9182391)(5.0,-1.9032391)(4.305,-1.9082391)(3.64,-1.873239)(3.175,-1.8282391)(2.675,-1.7932391)(3.055,-1.2782391)(3.58,-0.5332391)(3.87,-0.1532391)(4.4,0.6017609)(4.945,1.3017609)
\pspolygon[linecolor=colour1, linewidth=0.02, fillstyle=solid,fillcolor=colour1](4.05,0.1017609)(4.26,0.0717609)(4.475,0.056760903)(4.83,0.0467609)(5.11,0.0467609)(5.52,0.0517609)(5.96,0.056760903)(6.52,0.081760906)(6.815,0.0917609)(7.31,-0.5432391)(7.545,-0.8582391)(7.125,-0.9282391)(6.62,-0.9632391)(6.085,-0.9832391)(5.705,-0.9982391)(5.3,-0.9932391)(4.915,-1.0032392)(4.51,-0.9932391)(4.12,-0.9832391)(3.8,-0.9582391)(3.445,-0.9182391)(3.34,-0.8832391)
\psbezier[linecolor=black, linewidth=0.02, fillstyle=solid,fillcolor=colour1](5.075,1.2367609)(5.2393365,1.4324476)(5.5816946,1.506445)(5.755,1.1132314906400802)(5.9283056,0.72001785)(5.737054,0.3562439)(5.218158,0.6067609)(4.6992617,0.8572779)(4.9106636,1.0410742)(5.075,1.2367609)
\rput[bl](5.185,1.8567609){\normalsize{$z$}}
\rput[bl](6.945,-1.5332391){\normalsize{$K$}}
\rput[bl](0.0,-1.8832392){\normalsize{$\scrN=\scrN_0$}}
\psbezier[linecolor=black, linewidth=0.04](5.285,1.6967609)(4.375,0.7667609)(3.43,-0.9382391)(2.46,-1.9932390975952148)
\psbezier[linecolor=black, linewidth=0.04](5.27,1.6917609)(6.3,0.8617609)(7.115,-0.1532391)(8.165,-1.778239097595215)
\pscircle[linecolor=black, linewidth=0.04, fillstyle=solid,fillcolor=black, dimen=outer](5.215,0.8767609){0.04}
\rput[bl](5.365,0.7167609){\normalsize{$x_0$}}
\rput[bl](5.275,1.0767609){\normalsize{$\Omega$}}
\psbezier[linecolor=black, linewidth=0.03](1.555,-1.7682391)(3.146376,-1.7806783)(3.6400008,-1.9044296)(4.735,-1.913239097595215)(5.8299994,-1.9220486)(7.538506,-1.8478837)(9.115,-1.6832391)
\psbezier[linecolor=black, linewidth=0.03](1.4959846,-0.009933323)(2.8149264,-0.03185253)(3.3458576,-0.095716484)(4.414715,-0.1404833065456296)(5.483572,-0.18525013)(7.04329,-0.1535878)(8.106405,0.0034194803)
\psbezier[linecolor=black, linewidth=0.03](1.495,-0.6432391)(3.086376,-0.6556783)(3.9850006,-0.78442955)(4.985,-0.7832390975952148)(5.984999,-0.78204864)(6.7335057,-0.7878837)(8.015,-0.6232391)
\rput[bl](0.295,-0.7882391){\normalsize{$\scrN_{t_{j-1}}$}}
\rput[bl](0.285,-0.1432391){\normalsize{$\scrN_{t_j}$}}
\psbezier[linecolor=black, linewidth=0.02](4.03,0.1167609)(4.84,-0.0032390975)(6.35,0.0867609)(6.815,0.10676090240478515)
\psbezier[linecolor=black, linewidth=0.02](3.33,-0.8982391)(4.02,-1.0732391)(6.965,-1.0332391)(7.545,-0.8682390975952149)
\rput[bl](6.42,-0.6282391){\normalsize{$V_j$}}
\end{pspicture}
}
\caption{Covering by convex neighborhoods.}
\label{figcone}
\end{figure}

Let~$x_0 \in \scrM$. We want to show that~$T^\varepsilon$ is of regularized
Hadamard form at~$x_0$.
Without loss of generality we can assume that~$x_0$ lies in the future
of the Cauchy surface~$\scrN$. Closely following the construction in~\cite{fulling+sweeny+wald},
we choose a point~$z$ in the future of~$x_0$
and let~$\Omega$ be a convex neighborhood of~$x_0$ contained in the causal past~$J^\wedge(z)$
of~$z$ (see Figure~\ref{figcone}).

\begin{figure}
%
\psscalebox{1.0 1.0} 
{
\begin{pspicture}(0,-1.7116609)(13.311621,1.7116609)
\definecolor{colour0}{rgb}{0.8,0.8,0.8}
\definecolor{colour1}{rgb}{0.6,0.6,0.6}
\psbezier[linecolor=black, linewidth=0.02, fillstyle=solid,fillcolor=colour0](9.460123,1.0929537)(10.146468,1.8135777)(11.576317,2.0860765)(12.300122,0.6380516830145265)(13.023929,-0.80997306)(12.225172,-2.1495852)(10.058018,-1.2270464)(7.890863,-0.30450755)(8.773777,0.37232962)(9.460123,1.0929537)
\pspolygon[linecolor=colour1, linewidth=0.01, fillstyle=solid,fillcolor=colour1](8.075123,0.57795364)(8.335123,0.57795364)(8.700123,0.5329536)(9.325123,0.43795365)(9.740123,0.40795365)(10.195123,0.44795364)(10.580123,0.5229536)(10.930122,0.5879536)(11.4351225,0.67295367)(11.860123,0.74295366)(12.235123,0.7879536)(12.780123,0.86295366)(13.195123,0.9129536)(13.180122,0.7129536)(12.925122,0.68795365)(12.440123,0.63295364)(11.940123,0.54795367)(11.485123,0.48795363)(11.115123,0.39295363)(10.705123,0.31295365)(10.260122,0.19295365)(9.890122,0.16295364)(9.480123,0.19295365)(8.995123,0.26295364)(8.715122,0.30295363)(8.330123,0.33295363)(8.080123,0.34795365)
\psbezier[linecolor=black, linewidth=0.02, fillstyle=solid,fillcolor=colour0](1.4601227,1.0929537)(2.1464682,1.8135777)(3.576317,2.0860765)(4.3001227,0.6380516830145265)(5.023928,-0.80997306)(4.2251716,-2.1495852)(2.0580175,-1.2270464)(-0.10913688,-0.30450755)(0.7737771,0.37232962)(1.4601227,1.0929537)
\pspolygon[linecolor=colour1, linewidth=0.01, fillstyle=solid,fillcolor=colour1](0.08012268,0.51295364)(0.23512268,0.51795363)(0.5801227,0.50295365)(1.1451226,0.46295366)(1.5951227,0.41795364)(2.1201227,0.37795365)(2.5951226,0.38295364)(2.9901228,0.40295365)(3.4901228,0.42795363)(4.075123,0.46795365)(4.4451227,0.50795364)(4.9151225,0.55795366)(5.2901225,0.61795366)(5.2951226,0.41295364)(5.0001225,0.37295365)(4.4401226,0.30795366)(3.9401226,0.25795364)(3.4951227,0.22795364)(3.1151228,0.20295364)(2.7751226,0.19295365)(2.4351227,0.18295364)(1.9701227,0.19295365)(1.5301226,0.23295364)(1.0401226,0.28795364)(0.6801227,0.31795365)(0.33012268,0.33295363)(0.08012268,0.34795365)
\pscircle[linecolor=black, linewidth=0.02, dimen=outer](1.9826226,0.29545364){0.7125}
\pscircle[linecolor=black, linewidth=0.04, fillstyle=solid,fillcolor=black, dimen=outer](1.9701227,0.30295363){0.04}
\rput[bl](3.8951228,1.4079536){\normalsize{$\Omega$}}
\psbezier[linecolor=black, linewidth=0.02](0.08012268,-1.6220464)(1.1810348,-1.6344855)(1.4925239,-1.6282369)(2.2500434,-1.637046356201172)(3.0075626,-1.6458559)(4.219506,-1.7016909)(5.3101225,-1.5370463)
\psbezier[linecolor=black, linewidth=0.02](0.07012268,-1.0320463)(1.1710348,-1.0444856)(1.252524,-1.0982368)(2.0100434,-1.1070463562011719)(2.7675626,-1.1158559)(4.2095056,-1.111691)(5.3001227,-0.94704634)
\psbezier[linecolor=black, linewidth=0.02](0.09012268,1.2179537)(1.1910348,1.2055144)(1.5925239,1.1417632)(2.3500433,1.1329536437988281)(3.1075628,1.1241441)(4.2295055,1.138309)(5.3201227,1.3029536)
\psbezier[linecolor=black, linewidth=0.02](0.07012268,0.5279536)(1.1710348,0.5155145)(1.512524,0.39176318)(2.2700434,0.3829536437988281)(3.0275626,0.3741441)(4.2095056,0.44830903)(5.3001227,0.61295366)
\psbezier[linecolor=black, linewidth=0.02](0.050122682,-0.49204636)(1.1510348,-0.50448555)(1.7425239,-0.55823684)(2.5800433,-0.6070463562011719)(3.4175627,-0.6558559)(4.1895056,-0.571691)(5.2801228,-0.40704635)
\psbezier[linecolor=black, linewidth=0.02](0.0,-0.0070463563)(1.1010349,-0.019485522)(1.392524,-0.123236835)(2.0600433,-0.16704635620117186)(2.7275627,-0.21085589)(4.139506,-0.16169095)(5.2301226,0.002953644)
\psbezier[linecolor=black, linewidth=0.02](0.08012268,0.32795364)(1.1810348,0.31551448)(1.522524,0.19176316)(2.2800434,0.18295364379882811)(3.0375626,0.17414412)(4.219506,0.24830905)(5.3101225,0.41295364)
\pscircle[linecolor=black, linewidth=0.04, fillstyle=solid,fillcolor=black, dimen=outer](3.2601228,0.30295363){0.04}
\rput[bl](1.8951226,0.43795365){\normalsize{$x$}}
\rput[bl](3.3151226,-0.072046354){\normalsize{$y$}}
\rput[bl](2.6301227,0.6029537){\normalsize{$B_1(x)$}}
\rput[bl](5.3801227,0.09795365){\normalsize{$\scrN_{t(x)-\delta}$}}
\rput[bl](5.3751225,0.51795363){\normalsize{$\scrN_{t(x)+\delta}$}}
\pscircle[linecolor=black, linewidth=0.02, dimen=outer](9.982623,0.29545364){0.7125}
\pscircle[linecolor=black, linewidth=0.04, fillstyle=solid,fillcolor=black, dimen=outer](9.970122,0.30295363){0.04}
\pscircle[linecolor=black, linewidth=0.04, fillstyle=solid,fillcolor=black, dimen=outer](11.260122,0.30295363){0.04}
\rput[bl](9.895123,0.43795365){\normalsize{$x$}}
\rput[bl](11.345122,0.092953645){\normalsize{$y$}}
\psbezier[linecolor=black, linewidth=0.02](8.080123,-1.6220464)(9.181035,-1.6344855)(9.492524,-1.6282369)(10.250043,-1.637046356201172)(11.007563,-1.6458559)(12.219505,-1.7016909)(13.3101225,-1.5370463)
\psbezier[linecolor=black, linewidth=0.02](8.070123,-1.0320463)(9.171035,-1.0444856)(9.252524,-1.0982368)(10.010043,-1.1070463562011719)(10.767563,-1.1158559)(12.209506,-1.111691)(13.300122,-0.94704634)
\psbezier[linecolor=black, linewidth=0.02](8.050122,-0.49204636)(9.151034,-0.50448555)(9.857524,-0.62323684)(10.580043,-0.6070463562011719)(11.302563,-0.5908559)(12.159506,-0.37669095)(13.280123,-0.40704635)
\psbezier[linecolor=black, linewidth=0.02](8.010122,-0.027046356)(9.10472,-0.04533925)(9.305053,-0.13909118)(10.058228,-0.15204635620117188)(10.811401,-0.16500154)(12.075762,0.16082923)(13.210123,0.09795365)
\psbezier[linecolor=black, linewidth=0.02](8.090122,1.2179537)(9.191035,1.2055144)(9.592524,1.1417632)(10.350043,1.1329536437988281)(11.107563,1.1241441)(12.204506,1.273309)(13.295123,1.4379536)
\psbezier[linecolor=black, linewidth=0.02](8.080123,0.5879536)(9.181035,0.5755145)(9.314964,0.28295365)(10.440043,0.5029536437988281)(11.565123,0.7229536)(12.114506,0.798309)(13.200123,0.91795367)
\psbezier[linecolor=black, linewidth=0.02](8.080123,0.32930878)(9.165141,0.31675336)(9.666122,0.054742698)(10.318283,0.21795364379882812)(10.970444,0.38116458)(12.332219,0.64761144)(13.200123,0.7229536)
\psbezier[linecolor=black, linewidth=0.04, arrowsize=0.05291667cm 2.0,arrowlength=2.4,arrowinset=0.0]{->}(5.680123,-0.46204636)(6.2151227,-1.0120463)(7.3951225,-0.61204636)(7.6051226,-0.5470463562011719)
\rput[bl](6.0601225,-1.2420464){\normalsize{deform}}
\end{pspicture}
}
\caption{Deformation of the foliation.}
\label{figdeform}
\end{figure}
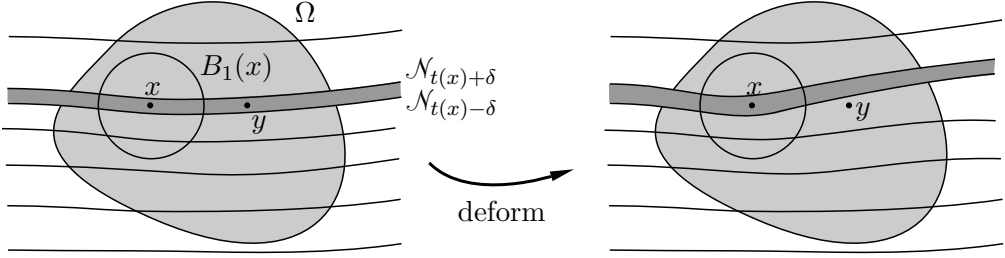

Let~$\mathfrak{t}$ be a Cauchy time function on~$\scrM$ with~$\mathfrak{t}^{-1}(0)=\scrN$.
We denote the corresponding Cauchy surfaces by~$\scrN_t := \mathfrak{t}^{-1}(t)$.
We first argue that there is~$\delta>0$ such that it suffices to consider points~$x,y \in \Omega$
whose time functions differ at least by~$\delta$,
\[ |\mathfrak{t}(x) - \mathfrak{t}(y)| \geq \delta \:. \]
Indeed, for any points~$x,y \in \Omega$ with~$d(x,y)>1$ this can be arranged by
slightly modifying the time function in a neighborhood of~$\Omega$ (see Figure~\ref{figdeform}).
Moreover, by symmetry in~$x$ and~$y$, it suffices to consider the case that~$x$ lies to the future of~$y$, i.e.
\beq \label{tcond}
\mathfrak{t}(x) - \mathfrak{t}(y) \geq \delta \:.
\eeq

Since~$\scrM$ is globally hyperbolic, the set~$K := J^\wedge(z) \cap J^\vee(\scrN)$
is compact. We choose~$\delta_1, \delta_2>0$ such that the set~$Q(x,\delta_1, \delta_2)$
defined by
\[ Q(x,\delta_1, \delta_2) = \big\{ y \in M \:\big|\: |\mathfrak{t}(x)-\mathfrak{t}(y)| < \delta_1 \text{ and }
\Gamma(x,y) > -\delta_2 \big\} \]
is a normal neighborhood for all~$x \in K$.
Let~$N$ be an integer with $N \delta_1 >2 t(z)$. Let~$t_j = j \,t(z)/N$ and let~$\scrN_j$
be the Cauchy surface~$t=t_j$. Finally, we let
\[ V_j = \Big\{ w \in K \;\Big|\; t_{j-1} - \frac{\delta_1}{4} < t(w) < t_j + \frac{\delta_1}{4} \Big\}\:. \]
Now we can proceed inductively. Thus, assuming that~$T^\varepsilon$ is of regularized
Hadamard form on~$\scrN_{t_{j-1}} \cap K$, it remains to show that it is also of regularized Hadamard form
in the set~$V_j$ which includes~$\scrN_{t_j} \cap K$.

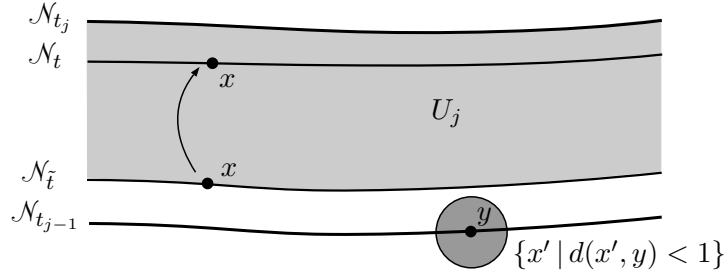
\begin{figure}
%
\psscalebox{1.0 1.0} 
{
\begin{pspicture}(0,-1.709834)(9.22,1.709834)
\definecolor{colour1}{rgb}{0.8,0.8,0.8}
\definecolor{colour0}{rgb}{0.6,0.6,0.6}
\pspolygon[linecolor=colour1, linewidth=0.02, fillstyle=solid,fillcolor=colour1](1.005,1.595166)(1.625,1.575166)(2.915,1.535166)(4.02,1.470166)(4.715,1.450166)(5.39,1.440166)(5.945,1.440166)(6.94,1.470166)(7.685,1.515166)(8.625,1.6151661)(8.625,-0.39983398)(7.48,-0.504834)(6.23,-0.589834)(5.215,-0.629834)(4.13,-0.639834)(3.05,-0.589834)(2.365,-0.53983396)(1.365,-0.509834)(1.01,-0.479834)
\pscircle[linecolor=black, linewidth=0.02, fillstyle=solid,fillcolor=colour0, dimen=outer](6.1125,-1.217334){0.4825}
\rput[bl](6.18,-1.1298339){\normalsize{$y$}}
\rput[bl](0.0,-1.2081673){\normalsize{$\scrN_{t_{j-1}}$}}
\psbezier[linecolor=black, linewidth=0.04](1.0266666,-1.099834)(2.6292691,-1.1122731)(3.1263764,-1.2360245)(4.2291007,-1.244833984375)(5.331825,-1.2536435)(7.0523844,-1.1794785)(8.64,-1.0148339)
\psbezier[linecolor=black, linewidth=0.04](1.0010853,1.5882659)(2.5255392,1.5664268)(3.1392016,1.5025951)(4.3746057,1.4578931270249231)(5.610009,1.4131912)(7.412752,1.4449482)(8.641505,1.60202)
\rput[bl](0.13833334,-0.64150065){\normalsize{$\scrN_{\tilde{t}}$}}
\rput[bl](0.235,1.430166){\normalsize{$\scrN_{t_j}$}}
\rput[bl](6.66,-1.709834){\normalsize{$\{x' \,|\, d(x',y)<1 \}$}}
\psbezier[linecolor=black, linewidth=0.03](0.99995553,-0.51984423)(2.6081867,-0.53017426)(3.1071992,-0.6532712)(4.2137966,-0.6606294558461469)(5.320394,-0.6679877)(7.0468802,-0.59155846)(8.639836,-0.4248246)
\pscircle[linecolor=black, linewidth=0.04, fillstyle=solid,fillcolor=black, dimen=outer](6.1025,-1.207334){0.0775}
\rput[bl](5.575,0.18516602){\normalsize{$U_j$}}
\pscircle[linecolor=black, linewidth=0.04, fillstyle=solid,fillcolor=black, dimen=outer](2.6675,1.032666){0.0775}
\pscircle[linecolor=black, linewidth=0.04, fillstyle=solid,fillcolor=black, dimen=outer](2.6025,-0.567334){0.0775}
\rput[bl](2.78,-0.46983397){\normalsize{$x$}}
\rput[bl](2.7566667,0.7218327){\normalsize{$x$}}
\psbezier[linecolor=black, linewidth=0.02, arrowsize=0.05291667cm 2.0,arrowlength=1.4,arrowinset=0.0]{->}(2.4433334,-0.41816732)(2.2077777,-0.058213137)(2.0655556,0.42132103)(2.4966667,0.9834993489583335)
\psbezier[linecolor=black, linewidth=0.03](1.0066222,1.0534891)(2.6148534,1.0431591)(3.1205325,1.0133954)(4.22713,1.0060372108205176)(5.3337274,0.998679)(7.053547,0.98177487)(8.646503,1.1485088)
\rput[bl](0.23166667,0.9718327){\normalsize{$\scrN_t$}}
\end{pspicture}
}
\caption{Cauchy development in~$x$.}
\label{figcone2}
\end{figure}

By the induction hypothesis, every point~$y \in \scrN_{t_{j-1}} \cap K$ has a convex neighborhood
where~$T^\varepsilon$ is of regularized Hadamard form.
We cover~$\scrN_{t_{j-1}} \cap K$ by a finite number of such convex neighborhoods
and denote their union by~$U$. We choose~$\tilde{t} > t_{j-1}$ such that~$\scrN_{\tilde{t}} \cap K
\subset U$ (see Figure~\ref{figcone2}).
Next we choose the distance function~$d$ such that the
corresponding distance of~$\scrN_{t_{j-1}} \cap K$ and~$\scrN_{\tilde{t}} \cap K$ is bigger than one.
Let~$y \in \scrN_{j-1} \cap K$. By extending~$R^\varepsilon(.,y)$ smoothly
to zero, the regularized Hadamard expansion is well-defined for all~$x \in \scrN_{\tilde{t}} \cap K$.
Moreover, choosing~$N$, we know by hypothesis that
\[ \big( T^\varepsilon - R^\varepsilon_N\big)(.,y) = W^{s,2}_\text{loc}
\big( \scrN_{\tilde{t}} \cap K \big) \: \O\big(\varepsilon^2 \big)
+ W^{N,2}_\text{loc} \big(\scrN_{\tilde{t}} \cap K \big) \:. \]

We now consider the transport equations for the functions~$A_{n,k}$, $B_{n,k}$
and~$\fa_{n,k}$, $\fb_{n,k}$. According to Lemma~\ref{lemmastructure},
these transport equations all have the general form~\eqref{genform}.
Since~$\nabla \Gamma$ is non-zero in a convex neighborhood away from the diagonal,
these transport equations all have unique solutions away from the diagonal.
Therefore, we can extend the regularized Hadamard expansion~$R^\varepsilon_N(.,y)$
to the set
\[ U_j := V_j \cap K \cap J^\vee \big(\scrN_{\tilde{t}} \big) \:. \]
According to~\eqref{sval}, we know furthermore that
for all~$t \in [\tilde{t}, t_j]$,
\[ \big( \Box_x + \mu(x) \big) \, \big( T^\varepsilon - R^\varepsilon_N \big)(.,y)
\in W^{s-1,2}_\text{loc}\big( \scrN_t \cap K \big) + W^{N,2}\big( \scrN_t \cap K \big) \:. \]
Solving the Cauchy problem for~$T^\varepsilon - R^\varepsilon_N$ in~$U_j$,
we can apply Lemma~\ref{lemmacauchy} to conclude that
for all~$t \in [\tilde{t}, t_j]$,
\[ \big( T^\varepsilon - R^\varepsilon_N\big)(.,y) = W^{s,2}_\text{loc}
\big( \scrN_t \cap K \big) \: \O\big(\varepsilon^2 \big) + W^{N,2}_\text{loc}\big(\scrN_t \cap K\big) \:. \]
We note that, since the set~$\scrN_{t_{j-1}} \cap K$ is compact,
this estimate is uniform in~$y$.

In order to translate~$y$ to the future, we proceed as follows.
Fix~$x$ for example on the surface~$\scrN_{t_j}$.
Exchanging the roles of~$x$ and~$y$ in the above argument, we can
solve the Cauchy problem in the variable~$y$ to the future up to
any Cauchy surface which has the property that that its time distance
from~$\scrN_{t_j}$ is at least~$\delta$.
With this procedure, we obtain the desired regularized Hadamard expansion
for all~$x$ and~$y$ in~$V_j$, under the constraint that~$x$ lies to the future of~$y$
and that~\eqref{tcond} holds. This concludes the proof.
\QED

\section{An Explicit Example and Glueing Constructions} \label{secex}

\begin{Example} \label{exexplicit}
{\em{ We let~$\scrM = \R^{1,3}$ be four-dimensional Minkowski space.
Let~$T(x,y)$ be the distribution
\[ T(x,y) = \int \frac{d^4k}{(2 \pi)^4}\: \delta(k^2-m^2)\: \Theta(-k^0) \: e^{-ik(x-y)} \]
(where~$k=(k^0, \vec{k})$ and~$k^2 = (k^0)^2 - |\vec{k}|^2$ is the Minkowski inner product).
By direct computation, one sees that this distribution is a bi-solution of the Klein-Gordon equation
\[ \big( \Box - m^2 \big) \phi = 0 \]
(where~$\Box = \partial_t^2 - \Box_{\R^3}$ is the wave operator).
Computing the integrals in terms of Bessel functions and 
expanding in powers of~$m^2$ (for details see for example~\cite[Section~1.2]{cfs})
one sees that this distribution is indeed of Hadamard form~\eqref{Teps}.
The simplest method to regularize it is to insert a convergence-generating factor~$e^{\varepsilon k^0}$
into the integrand,
\beq \label{Tepsex}
T^\varepsilon(x,y) := \int \frac{d^4k}{(2 \pi)^4}\: \delta(k^2-m^2)\: \Theta(-k^0)\: e^{\varepsilon k^0} \:
e^{-ik(x-y)}\:.
\eeq
The effect of the convergence-generating factor
can be described conveniently in position space. Namely, introducing the short notations
\[ \omega = k^0 \qquad \text{and} \qquad y-x = (t, \vec{\xi}) \:, \]
one can combine the exponential with the phase factor of the Fourier transform,
\[ \exp(\varepsilon k^0)\: e^{i k \xi} = e^{i \omega (t - i \varepsilon) - i \vec{k} \vec{\xi}} \:. \]
This shows that the regularization amounts to the replacement
\[ 
t \rightarrow t - i \varepsilon \:. \]
This simple replacement rule motivates the name {\em{$i \varepsilon$-regularization}}
(see~\cite[Section~2.4]{cfs}).
According to~\eqref{Gammareg}, this replacement rules gives rise to
the regularized Hadamard expansion with~$\f\equiv 2$, up to
errors of the form~\eqref{fehler}. Even more, writing the replacement rule as
\[ \Gamma \rightarrow \Gamma_\varepsilon - \varepsilon^2 \]
and expanding in powers of~$\varepsilon^2$, one obtains an expansion of the
form considered in Section~\ref{secprop}. More precisely, truncating in~$n$ and~$k$, we
obtain the expansion~$R^\varepsilon_N$ in~\eqref{RepsN}.
Therefore, $T^\varepsilon(x,y)$ is of {\em{regularized Hadamard form
on~$\scrM$}} to the order~$\O^s(\varepsilon^2)$ for any~$s \in \N$
(see Definition~\ref{defregHad}).
\QEDrem }}
\end{Example}

Combining this explicit example with the result on the propagation of
regularized singularities of Theorem~\ref{thmpropagate}, we can adapt
the well-known glueing constructions for Hadamard states in~\cite{fulling+narcowich+wald} to the case with
regularization:
\begin{Thm} \label{thmglue}
Let~$(\scrM, g)$ be a four-dimensional globally hyperbolic space-time
whose Cauchy surfaces are diffeomorphic to~$\R^3$. Then for any~$s \in \N$,
there is a family~$T^\varepsilon$ of distributions which are 
of regularized Hadamard form in all of~$\scrM$ to the order~$\O^s(\varepsilon^2)$.
\end{Thm}
\Proof We choose a global foliation by Cauchy surfaces~$(\scrN_t)_{t \in \R}$.
We proceed in two steps. In the first step, we take the future of the Cauchy surface~$\scrN_1$
and glue it together with the subset~$\R^- \times \R^3$ of Minkowski space.
More precisely, we apply the construction in~\cite[Proposition~C.1]{fulling+narcowich+wald}
to obtain a globally hyperbolic space-time~$\tscrM$ with foliation by Cauchy
surfaces~$(\tscrN_t)_{t \in \R}$ such that the submanifold~$\cup_{t > 1} \tscrN_t$
is isometric to~$\cup_{t > 1} \scrN_t$, whereas the submanifold~$\cup_{t <0} \tscrN_t$
is isometric to the subset~$\R^- \times \R^3$ of Minkowski space.
We let~$\tilde{T}^\varepsilon$ be the family of bi-distributions on~$\tscrM \times \tscrM$
which for negative times coincides with the explicit example in Minkowski space~\eqref{Tepsex}.
Applying Theorem~\ref{thmpropagate}, it follows that~$\tilde{T}^\varepsilon$ is of 
regularized Hadamard form in all of~$\tscrM$.

In the second step, we let~$T^\varepsilon$ be the family of bi-distributions on~$\scrM \times \scrM$
which for for times~$t>1$ coincides with~$\tilde{T}^\varepsilon$.
Again applying Theorem~\ref{thmpropagate}, it follows that~$T^\varepsilon$ is of 
regularized Hadamard form in all of~$\scrM$.
\QED

\section{A Dirac Field} \label{secdirac}
We now explain how the above methods and results extend to a Dirac field,
also relying on constructions in~\cite{lqg, dgc}.
Thus we consider a family~$P^\varepsilon(x,y)$ of solutions of the Dirac equation,
\[ (\Dir - m)\, P^\varepsilon(x,y) = 0 \:. \]
Similar as in~\cite[Lemma~5.6]{lqg}, it is convenient to write this family as
\beq \label{PT}
P^\varepsilon(x,y) = (\Dir + m) \,T^\varepsilon(x,y)
\eeq
($T^\varepsilon(x,y)$ can be obtained for example
by solving this inhomogeneous hyperbolic equation for
any given initial data). Then~$T^\varepsilon(x,y)$ satisfies the equation
\[ 0 = (\Dir - m)(\Dir + m) \,T^\varepsilon(x,y) = \big(\Dir^2 - m^2\big) \,T^\varepsilon(x,y) 
= \Big(-\Box^S +\frac{\s}{4}-m^2\Big) \,T^\varepsilon(x,y) \:, \]
where we used the Lichnerowicz-Weitzenb\"ock identity (see for example~\cite[Theorem~II.8.8]{lawson+michelsohn}
or~\cite{schroedinger, lichnerowicz}),
and~$\s$ denotes scalar curvature. Choosing
\[ \mu(x) = m^2 - \frac{\s(x)}{4} \:, \]
we obtain the equation
\[ \big( \Box^S_x +\mu(x) \big)\,T^\varepsilon(x,y) = 0 \:, \]
which coincides with~\eqref{Teq}, but now taking values in the spinor bundle~$S\scrM$.
In analogy to~\eqref{reghadamard}, for the regularized Hadamard expansion of~$T^\varepsilon$ we make the ansatz
\beq \label{reghadamardspinor}
T^\varepsilon = \frac{1}{ \Gamma_{[-1]}}\: X_{-1} + \sum_{n=0}^{\infty} \Gamma_{[n]}^n\:\log \Gamma_{[n]} \:X_n
+ \sum_{n=0}^{\infty} \Gamma_{\{n\}}^n\:Y_n \:,
\eeq
where the coefficients now are linear mappings between spinor spaces:
\begin{align*}
X_\bullet(x,y), Y_\bullet(x,y)  &:\: S_y\scrM \rightarrow S_x\scrM \:,
\end{align*}
whereas~$\Gamma_\bullet$ are again complex-valued functions.
Exactly as explained for the Klein-Gordon equation in Section~\ref{secscal},
we again work with the ansatz~\eqref{Gammaf}--\eqref{YB}.
Finally, the regularized Hadamard expansion for~$P^\varepsilon(x,y)$ is obtained from~\eqref{reghadamardspinor}
by differentiation~\eqref{PT}.

In this bundle setting, the derivation of the transport equations is more subtle because
one must carefully keep track of the order of multiplication of the linear operators
acting on spinors. We wrote the computations in Section~\ref{sectransport} and Appendix~\ref{appendixA}
in such a way that they again apply without changes. In particular, the
above transport equations~\eqref{Xm1}--\eqref{Ynp1}
and~\eqref{Am1}--\eqref{Bn2} again hold as spinorial equations, whereas~\eqref{fm1}
is complex-valued. These
differential equations can be solved
exactly as described in Section~\ref{secsolve}, again with the boundary conditions~\eqref{fzero}.
The solutions of the usual transport equations~\eqref{Xm1}--\eqref{Ynp1}
are computed more explicitly for example in~\cite[Appendix~A]{lqg}.

\appendix
\section{Derivation of the Transport Equations} \label{appendixA}
We substitute the regularized Hadamard expansion~\eqref{reghadamard}
into the Klein-Gordon equation~\eqref{Teq} and compute the Laplacian
term by term, omitting error terms of the order~\eqref{fehler}.
From now on all derivatives act on the variable~$x$. 
We remark that the identity
\[ 
|\nabla \Gamma|^2 \equiv (\nabla_j \Gamma)(\nabla^j \Gamma) = 4\, \Gamma \]
holds (see~\cite[Lemma~1.3.19]{baer+ginoux} or~\cite[Lemma~B.1]{dgc}). As a consequence,
\begin{align}
\left| \nabla \Gamma_\bullet \right|^2 &= \left|\nabla \Gamma\right|^2 - \varepsilon^2 \left|\nabla \f_\bullet\right|^2 
+ 2i\varepsilon \:\langle \nabla \Gamma, \nabla \f_\bullet \rangle \nonumber \\
&= 4 \Gamma - \varepsilon^2 \left|\nabla \f_\bullet\right|^2 
+ 2i\varepsilon \:\langle \nabla \Gamma, \nabla \f_\bullet \rangle \nonumber \\
&= 4\Gamma_\bullet- 4i\varepsilon \f_\bullet - \varepsilon^2 |\nabla \f_\bullet|^2 + 2 i \varepsilon \:\langle \nabla \Gamma,
\nabla \f_\bullet \rangle \label{nabGam} \:,
\end{align}
(where the bullets again stand for any subscript~$[\cdot]$ or~$\{\cdot\}$). Hence
\begin{align*}
\Box \Big(\frac{1}{\Gamma_{[-1]}} \: X_{-1}\Big)
&= \frac{1}{\Gamma_{[-1]}}\: \Box X_{-1} - \frac{2}{\Gamma^2_{[-1]}}\, \langle \nabla \Gamma_{[-1]}, \nabla X_{-1}
\rangle \\
&\qquad + \frac{2}{\Gamma^3_{[-1]}}\, |\nabla \Gamma_{[-1]}|^2\: X_{-1} - \frac{1}{\Gamma^2_{[-1]}}\,\Box\, \Gamma_{[-1]}\: X_{-1} \notag \\
&\!\!\overset{\eqref{nabGam}}{=} \frac{i \varepsilon}{\Gamma^3_{[-1]}} \,\Big(4\, \langle \nabla \Gamma,
\nabla \f_{[-1]} \rangle - 8\f_{[-1]}\Big)\:X_{-1} \notag \\
&\qquad - \frac{2 \varepsilon^2}{\Gamma^3_{[-1]}}\, \big|\nabla \f_{[-1]} \big|^2 \,X_{-1} \notag \\
&\qquad + \frac{1}{\Gamma^2_{[-1]}}\Big( -2 \langle \nabla \Gamma, \nabla X_{-1} \rangle + 8\,X_{-1} - (\Box\, \Gamma) \, X_{-1} \Big) \notag \\
&\qquad + \frac{i\varepsilon}{\Gamma^2_{[-1]}}\Big(-2 \langle \nabla \f_{[-1]}, \nabla X_{-1} \rangle -(\Box\, \f_{[-1]})
\, X_{-1}\Big) \notag \\
&\qquad + \frac{1}{\Gamma_{[-1]}}\Box X_{-1} \notag \:.
\end{align*}
Next, for all~$n=0,1,2,\ldots$,
\begin{align*}
\Box\, \Gamma^n_{\{n\}}
&= n \,\Gamma^{n-1}_{\{n\}}\, \Box\, \Gamma_{\{n\}} + n(n-1) \,\Gamma^{n-2}_{\{n\}}
 \,|\nabla\Gamma_{\{n\}}|^2 \notag \\
&\!\!\overset{\eqref{nabGam}}{=} n \,\Gamma^{n-1}_{\{n\}}\, \Box\, \Gamma_{\{n\}} \notag \\
&\qquad + n(n-1) \,\Gamma^{n-2}_{\{n\}} \Big( 
4\Gamma_{\{n\}}- 4i\varepsilon \f_{\{n\}} - \varepsilon^2 |\nabla \f_{\{n\}}|^2 + 2 i \varepsilon \:\langle \nabla \Gamma,
\nabla \f_{\{n\}} \rangle \Big) \notag \\
\Box \big( \Gamma^n_{\{n\}}\:Y_n \big) &=
\Gamma^n_{\{n\}}\:(\Box Y_n) + 2n\,\Gamma^{n-1}_{\{n\}}\: \langle \nabla\Gamma_{\{n\}} , \nabla Y_n\rangle +
(\Box\, \Gamma^n_{\{n\}})\:Y_n \notag \\
&= \Gamma^n_{\{n\}}\:(\Box Y_n) + 2n\,\Gamma^{n-1}_{\{n\}}\: \langle \nabla \Gamma_{\{n\}}, \nabla Y_n \rangle \notag \\
&\qquad + n \,\Gamma^{n-1}_{\{n\}}\, \Box\, \Gamma_{\{n\}}\:Y_n \notag \\
&\qquad + n(n-1) \,\Gamma^{n-2}_{\{n\}} \Big( 
4\Gamma_{\{n\}}- 4i\varepsilon \f_{\{n\}} - \varepsilon^2 |\nabla \f_{\{n\}}|^2 + 2 i \varepsilon \:\langle \nabla \Gamma,
\nabla \f_{\{n\}} \rangle \Big)\:Y_n \notag \\
&= i \varepsilon \,\Gamma^{n-2}_{\{n\}} \:2n(n-1)\:
\Big(-2\f_{\{n\}} + \langle \nabla \Gamma, \nabla \f_{\{n\}}\rangle\Big)\:Y_n
\notag \\
&\qquad + \Gamma^{n-1}_{\{n\}}\: n \,\Big( 2 \,\langle \nabla \Gamma, \nabla Y_n \rangle
+ (\Box\, \Gamma)\:Y_n\:  + 4(n-1) \,Y_n\Big) \notag \\
&\qquad + i \varepsilon \,\Gamma^{n-1}_{\{n\}}\: n\, \Big( 2 \,\langle \nabla \f_{\{n\}}, \nabla Y_n \rangle
+ \Box\, \f_{\{n\}}\:Y_n \Big) \notag \\
&\qquad + \Gamma^n_{\{n\}}\: (\Box Y_n) \notag \\
&\qquad - \varepsilon^2 \: \Gamma^{n-2}_{\{n\}}\: n(n-1)\:|\nabla \f_{\{n\}}|^2\:Y_n \notag
\end{align*}
We  now compute the terms involving logarithms,
\begin{align}
\Box \big(\Gamma^n_{[n]} \log \Gamma_{[n]} \big) &= 
\nabla\Big( \big( n \Gamma^{n-1}_{[n]}\: \log \Gamma_{[n]} + \Gamma^{n-1}_{[n]} 
\big) \nabla \Gamma_{[n]} \Big) \notag \\
&= \big( n \Gamma^{n-1}_{[n]}\: \log \Gamma_{[n]} + \Gamma^{n-1}_{[n]} 
\big) \Box\, \Gamma_{[n]} \notag \\
&\qquad + \Big( n (n-1) \:\Gamma^{n-2}_{[n]}\: \log \Gamma_{[n]} + (2n-1)\: \Gamma^{n-2}_{[n]} 
\Big) \:|\nabla \Gamma_{[n]}|^2 \notag \\
&\!\!\overset{\eqref{nabGam}}{=}
\big( n \Gamma^{n-1}_{[n]}\: \log \Gamma_{[n]} + \Gamma^{n-1}_{[n]} 
\big) \Box\, \Gamma_{[n]} \notag \\
&\qquad + 4 n (n-1) \:\Gamma^{n-1}_{[n]}\: \log \Gamma_{[n]} + 4 (2n-1)\: \Gamma^{n-1}_{[n]} 
\notag \\
&\qquad -4i \varepsilon \f_{[n]}\: n (n-1) \:\Gamma^{n-2}_{[n]}\: \log \Gamma_{[n]} -4i \varepsilon \f_{[n]}\:
(2n-1)\: \Gamma^{n-2}_{[n]} \notag \\
&\qquad -\varepsilon^2\, |\nabla \f_{[n]}|^2\:n (n-1) \:\Gamma^{n-2}_{[n]}\: \log \Gamma_{[n]}
- \varepsilon^2\, |\nabla \f_{[n]}|^2 \:(2n-1)\: \Gamma^{n-2}_{[n]} \notag \\
&\qquad +2 i \varepsilon \,\langle \nabla \Gamma,
\nabla \f_{[n]} \rangle\: n (n-1) \:\Gamma^{n-2}_{[n]}\: \log \Gamma_{[n]} \notag \\
&\qquad +2 i \varepsilon \:\langle \nabla \Gamma,
\nabla \f_{[n]} \rangle\: (2n-1)\: \Gamma^{n-2}_{[n]}  \notag \\
\Box \big( \Gamma^n_{[n]} \,\log \Gamma_{[n]} \:X_n \big)
&= \Gamma^n_{[n]} \:\log \Gamma_{[n]}\: (\Box X_n)
+ 2 \Gamma^{n-1}_{[n]}\langle \nabla \Gamma_{[n]}, \nabla X_n \rangle \notag \\
&\qquad + 2 n \,\Gamma^{n-1}_{[n]}\: \log \Gamma_{[n]} \:
\langle \nabla \Gamma_{[n]}, \nabla X_n \rangle 
+ \Box \big(\Gamma^n_{[n]}\log \Gamma_{[n]} \big) \: X_n \notag \\
&= \Gamma^n_{[n]} \:\log \Gamma_{[n]} \:(\Box X_n) \notag \\
&\qquad + 2 \Gamma^{n-1}_{[n]} \:
\Big(\langle \nabla \Gamma, \nabla X_n \rangle + i\varepsilon \,\langle \nabla \f_{[n]}, \nabla X_n\rangle\Big) \notag \\
&\qquad + 2 n \,\Gamma^{n-1}_{[n]} \log \Gamma_{[n]}\:
\Big(\langle \nabla \Gamma , \nabla X_n \rangle + i\varepsilon \,\langle \nabla \f_{[n]}, \nabla X_n \rangle\Big)\notag \\
&\qquad + \Box \big(\Gamma^n_{[n]}\log \Gamma_{[n]} \big)\: X_n\notag \\
&= \Gamma^n_{[n]} \:\log \Gamma_{[n]} \: (\Box X_n) \notag \\
&\qquad + \Gamma^{n-1}_{[n]}\Big( 2 \,\langle \nabla \Gamma, \nabla X_n \rangle + 4(2n-1) \,X_n
+  \Box\, \Gamma \:X_n \Big)\notag \\
&\qquad + i \varepsilon \,\Gamma^{n-1}_{[n]} \:
\Big(2 \,\langle \nabla \f_{[n]}, \nabla X_n \rangle + \Box\, \f_{[n]} \:X_n \Big)\notag \\
&\qquad + \Gamma^{n-1}_{[n]}\:\log \Gamma_{[n]}\:n\,\Big(2\,\langle \nabla \Gamma, \nabla X_n \rangle + 4(n-1)\,X_n +\Box\, \Gamma\: X_n \Big)\notag \\
&\qquad + i \varepsilon \,\Gamma^{n-1}_{[n]} \:\log \Gamma_{[n]} \:n\, \Big(2
\,\langle \nabla \f_{[n]}, \nabla X_n \rangle + \Box\, \f_{[n]} \:X_n \Big) \notag \\
&\qquad + i \varepsilon \,\Gamma^{n-2}_{[n]} \:2(2n-1) \:
\Big(-2\f_{[n]} \:X_n + \langle \nabla \Gamma,\nabla \f_{[n]} \rangle \Big) \notag \\
&\qquad - \varepsilon^2\: \Gamma^{n-2}_{[n]} \:(2n-1)\: |\nabla \f_{[n]}|^2\: X_n\notag \\
&\qquad +  i \varepsilon \,\Gamma^{n-2}_{[n]}\:\log  \Gamma_{[n]} \:2n(n-1)
\:\Big(-2 \f_{[n]}+\langle \nabla \Gamma,\nabla \f_{[n]} \rangle\Big) \:X_n \notag \\
&\qquad - \varepsilon^2 \:\Gamma^{n-2}_{[n]}\: \log \Gamma_{[n]}\:n(n-1)\:|\nabla \f_{[n]}|^2 \:X_n\:. \notag \\
&= i \varepsilon \,\Gamma^{n-2}_{[n]}\:\log  \Gamma_{[n]} \:2n(n-1)
\:\Big(-2 \f_{[n]}+\langle \nabla \Gamma,\nabla \f_{[n]} \rangle\Big)\: X_n \notag \\
&\qquad + i \varepsilon \,\Gamma^{n-2}_{[n]} \:2(2n-1) \:
\Big(-2\f_{[n]} + \langle \nabla \Gamma,\nabla \f_{[n]} \rangle \Big) \:X_n \notag \\
&\qquad + \Gamma^{n-1}_{[n]}\:\log \Gamma_{[n]}\:n\,\Big(2\,\langle \nabla \Gamma, \nabla X_n \rangle + 4(n-1)\,X_n +\Box\, \Gamma\: X_n \Big)\notag \\
&\qquad + i \varepsilon \,\Gamma^{n-1}_{[n]} \:\log \Gamma_{[n]} \:n\, \Big(2
\,\langle \nabla \f_{[n]}, \nabla X_n \rangle + \Box\, \f_{[n]} \:X_n \Big) \notag \\ 
&\qquad + \Gamma^{n-1}_{[n]}\Big( 2 \,\langle \nabla \Gamma, \nabla X_n \rangle + 4(2n-1) \,X_n + \Box\, \Gamma
\:X_n \Big)\notag \\
&\qquad + i \varepsilon \,\Gamma^{n-1}_{[n]} \:
\Big(2 \,\langle \nabla \f_{[n]}, \nabla X_n \rangle + \Box\, \f_{[n]}\:X_n \Big)\notag \\
&\qquad +\Gamma^n_{[n]} \:\log \Gamma_{[n]} \: (\Box X_n)\: \notag \\
&\qquad - \varepsilon^2\: \Gamma^{n-2}_{[n]} \:\log \Gamma_{[n]} \:(2n-1)\: |\nabla \f_{[n]}|^2 \:X_n \notag \\
&\qquad - \varepsilon^2 \:\Gamma^{n-2}_{[n]} \:n(n-1)\: |\nabla \f_{[n]}|^2 \:X_n\:. \notag
\end{align}

Combining all the terms, we obtain
\beq \label{transans}
\begin{split}
\big(\Box_x + \mu(x) \big)\, T^\varepsilon(x,y)
&= \sum_{n=-2}^\infty \frak{L}_n(x,y) + i \varepsilon \sum_{n=-3}^\infty \frak{l}_n(x,y) \\
&\qquad  +\sum_{n=0}^\infty \frak{M}_n(x,y) + i \varepsilon \sum_{n=0}^\infty \frak{m}_n(x,y)  \;+\; \varepsilon^2 R(x,y) \:,
\end{split}
\eeq
where the terms are ordered according to their singularity on the light cone; namely
\[  \frak{l}_n,  \:{\frak{L}}_n \sim \left\{ \begin{array}{cl} \Gamma^n_\bullet & \text{if~$n<0$} \\
\Gamma_\bullet^n\: \log \Gamma_\bullet & \text{if~$n\geq0$} \end{array} \right.
\qquad \text{and} \qquad  \frak{m}_n, \:\frak{M}_n \sim \Gamma_\bullet^n \:. \]
The contributions~$\frak{L}_\bullet$ and~$\frak{M}_\bullet$ are computed by
(where always $n=0,1,2,\ldots$)
\begin{align*}
\frak{L}_{-2} &= - \Gamma_{[-1]}^{-2}\: \Big( 2 \,\langle \nabla \Gamma, \nabla A_{-1} \rangle - \big(8 - \Box\, \Gamma \big) \:A_{-1} \Big) \\
\frak{L}_{-1} &= \Gamma_{[0]}^{-1}\: \Big( 2 \,\langle \nabla \Gamma, \nabla A_0 \rangle - \big(4 - \Box\, \Gamma \big) \:A_0 \Big) \\
&\quad\, + \Gamma_{[-1]}^{-1}\: \big(\Box A_{-1} + \mu(x)\, A_{-1}\big)  \\
\frak{L}_n &= \Gamma_{[n+1]}^n \:\log \Gamma_{[n+1]}\: (n+1)\,\Big(2\,\langle \nabla \Gamma, \nabla A_{n+1} \rangle + \big( 4n + \Box\, \Gamma \big)\: A_{n+1} \Big)  \\
&\quad\, + \Gamma_{[n]}^n \:\log \Gamma_{[n]}\: \big(\Box A_n + \mu(x)\, A_n \big) \\
\frak{M}_n &= (n+1) \:\Gamma_{\{n+1\}}^n\,\Big( 2 \,\langle \nabla \Gamma, \nabla B_{n+1} \rangle
+ \big(4n + \Box\, \Gamma \big) \:B_{n+1} \Big)  \\
&\quad\, + \Gamma_{\{n\}}^n\:\big(\Box B_n + \mu(x)\, B_n \big) \\
&\quad\, + \Gamma_{[n+1]}^n\: \Big( 2 \,\langle \nabla \Gamma, \nabla A_{n+1} \rangle + \big(
4(2n+1) + \Box\, \Gamma \big) \: A_{n+1} \Big) \:.
\end{align*}
Solving the equation~$(\Box+\mu) T^\varepsilon=0$ gives rise to
the usual transport equations in the standard (i.e\ unregularized) Hadamard expansion
as given in Proposition~\ref{prphadamard1}.
Using these transport equations, the above functions simplify to
\begin{align}
\frak{L}_{-2} &= 0 \label{Lm2n} \\
\frak{L}_{-1} &= \big( \Gamma_{[-1]}^{-1} - \Gamma_{[0]}^{-1} \big)\:\big((\Box + \mu) A_{-1} \big) \\
\frak{L}_n &= \big( \Gamma_{[n]}^n \:\log \Gamma_{[n]} - \Gamma_{[n+1]}^n 
\:\log \Gamma_{[n+1]} \big) \: \big((\Box + \mu) A_n \big) \\
\frak{M}_n &= 4 (n+1)\: \big( \Gamma_{[n+1]}^n - \Gamma_{\{n+1\}}^n \big)\:A_{n+1} \\
&\quad\, + \big( \Gamma_{\{n\}}^n - \Gamma_{\{n+1\}}^n \big)\:\big( (\Box + \mu) B_n \big) \\
&\quad\, -\big( \Gamma_{[n+1]}^n - \Gamma_{\{n+1\}}^n \big)\:\frac{(\Box + \mu) A_n}{n+1} \:. \label{Mnn}
\end{align}
The formulas for~$\frak{L}_{-1}$, $\frak{L}_n$ and~$\frak{M}_n$
involve differences of expressions involving factors~$\Gamma_\bullet$ with different
lower indices. The next lemma gives relations between such expressions
\begin{Lemma} \label{lemmatransform}
For any~$n \in \Z$ and any subscripts~$a$ and~$b$,
\begin{align*}
\big( \Gamma_a^n - \Gamma_b^n \big)\:
\Big( 1 + \O \big( \varepsilon^2 / \Gamma^2 \big) \Big) &= i \varepsilon \: (\f_a - \f_b) \:n\: \Gamma^{n-1}_b \\
\Big( \Gamma_a^n \:\log \Gamma_a - \Gamma_b^n \:\log \Gamma_b \Big)
\Big( 1 + \O \big( \varepsilon^2 / \Gamma^2 \big) \Big)
&= i \varepsilon \: (\f_a - \f_b) \Big( n \,\Gamma^{n-1}_b \:\log \Gamma_b
+ \Gamma^{n-1}_b \Big) \:.
\end{align*}
\end{Lemma}
\Proof For any~$n \geq 0$, the binomial formula yields
\[ \Gamma_a^n - \Gamma_b^n = (\Gamma_b + i \varepsilon (\f_a - \f_b) \big)^n - \Gamma_b^n 
= n  \,i \varepsilon (\f_a - \f_b)\, \Gamma_b^{n-1} + \O\big( \varepsilon^2 /\Gamma \big)\: \Gamma_b^n \:. \]
For negative~$n$, we bring the terms on a common denominator,
\[ \Gamma_a^n - \Gamma_b^n = \Gamma_a^n \:\Gamma_b^n \:\big( \Gamma_b^{-n} - \Gamma_a^{-n} \big) \:, \]
making it possible to proceed just as in the case~$n>0$. If logarithms appear, we first organize the terms as
\[ \Gamma_a^n \: \log \Gamma_a - \Gamma_b^n \: \log \Gamma_b
= \big( \Gamma_a^n - \Gamma_b^n \big) \: \log \Gamma_a + \Gamma_b^n \: 
\big( \log \Gamma_a - \log \Gamma_b \big) \:. \]
The first summand can be treated as above. Rewriting the difference of the logarithms as
\[ \log \Gamma_a - \log \Gamma_b = \log \Big( \frac{\Gamma_a}{\Gamma_b} \Big)
= \log \Big( \frac{\Gamma_b +  i \varepsilon (\f_a - \f_b)}{\Gamma_b} \Big)
= \log \Big( 1 + i \varepsilon \:\frac{\f_a - \f_b}{\Gamma_b} \Big) \:, \]
and expanding the logarithm gives the result.
\QED
The error term in the above lemma requires a detailed explanation.
If~$y$ is not on the light cone centered at~$x$, the function~$\Gamma$ is non-zero,
so that the error term~$\O(\varepsilon^2/\Gamma^2) = \O(\varepsilon^2$)
is of higher order in~$\varepsilon$.
In particular, the error term becomes small as~$\varepsilon \searrow 0$.
In this sense, the transformations in Lemma~\ref{lemmatransform} are
well-defined away from the light cone.
On the light cone, however, when the function~$\Gamma$ vanishes, the
error terms in Lemma~\ref{lemmatransform} do not need to be small,
and the transformations are not sensible.
One may wonder whether the formulas in Lemma~\ref{lemmatransform}
can be given a mathematical meaning on the light cone.
An obvious idea would be to write the error term for example as
\beq \label{regerr}
\O \big( \varepsilon^2 / \Gamma_a^2 \big) \:.
\eeq
The identity
\[ \bigg| \frac{\varepsilon^2}{\Gamma_a^2} \bigg| =
\bigg| \frac{\varepsilon^2}{\Gamma^2 + \varepsilon^2 \f_a^2} \bigg| \leq 
\frac{1}{\f_a^2} \]
shows that the error term is uniformly bounded, provided that~$\f_a$ is non-zero.
But it is in general not small on the light cone. Therefore, working with
error terms of the form~\eqref{regerr} is not admissible.
In contrast, ``regularizing'' the error term~\eqref{fehler} accordingly gives
\beq \label{regerrgood}
\O \big( \varepsilon^2 / \Gamma_a \big) \qquad \text{with} \qquad
\bigg| \frac{\varepsilon^2}{\Gamma_a} \bigg|  \leq 
\frac{\varepsilon}{|\f_a|} \:.
\eeq
This error term is small even on the light cone, provided that~$\f_a$ is non-zero.
In what follows, we shall always work with error terms of the form~\eqref{regerrgood},
but errors like~\eqref{regerr} or the error terms in Lemma~\ref{lemmatransform}
are not good enough for us.

We conclude that, working with error terms of the form~\eqref{regerrgood},
we cannot transform the lower indices of the factors~$\Gamma_\bullet$.
Since the functions~$(\Box + \mu) A_\bullet$ and~$(\Box + \mu) B_\bullet$
in~\eqref{Lm2n}--\eqref{Mnn} in general do not vanish, we are led to choosing
all functions~$\f_\bullet$ equal,
\[ \f_{[n]} = \f_{\{n\}} = \f_{[-1]} \qquad \text{for all~$n=0,1,2,\ldots$}\:. \]
For ease in notation, we omit the lower indices of the function~$f$ and replace the
lower index of~$\Gamma_\bullet$ by an~$\varepsilon$,
\[ \f := \f_\bullet \qquad \text{and} \qquad \Gamma_\varepsilon := \Gamma_\bullet \:. \]
Then the conditions~\eqref{Lm2n}--\eqref{Mnn} are all satisfied.

We next consider the functions~$\frak{l}_\bullet$ and~$\frak{m}_\bullet$ in~\eqref{transans}.
The equation
\[ 0 = \frak{l}_{-3} = 4\,\Gamma_\varepsilon^{-3}\: \Big(\langle \nabla \Gamma, \nabla \f \rangle - 2\f \Big) \:X_{-1} \]
gives rise to the transport equation
\[ 
\langle \nabla \Gamma, \nabla \f \rangle = 2\f \:. \]
Using this transport equation, we obtain
\begin{align*}
\frak{l}_{-2} &= - \Gamma_\varepsilon^{-2}\: \Big( 2 \,\langle \nabla \Gamma, \nabla \fa_{-1} \rangle - \big(8 - \Box\, \Gamma \big) \:\fa_{-1} \Big) \\
&\quad\, -\Gamma_\varepsilon^{-2}\: \Big(2 \, \langle \nabla \f_, \nabla X_{-1} \rangle  + (\Box\, \f)\:X_{-1} \Big) \\
\frak{l}_{-1} &= \Gamma_\varepsilon^{-1}\: \Big( 2 \,\langle \nabla \Gamma, \nabla \fa_{0} \rangle - \big(4 - \Box\, \Gamma \big) \:\fa_{0} \Big) \\
&\quad\, + \Gamma_\varepsilon^{-1}\: \big(\Box\, \fa_{-1} + \mu(x)\, \fa_{-1}\big) \\
&\quad\, + \Gamma_\varepsilon^{-1}\:\Big(2 \,\langle \nabla \f_, \nabla X_0 \rangle + (\Box\, \f) \:X_0 \Big)
\end{align*}
\begin{align*}
\frak{l}_n &= \Gamma_\varepsilon^n \:\log \Gamma_\varepsilon\: (n+1)\,\Big(2\,\langle \nabla \Gamma, \nabla \fa_{n+1} \rangle + \big( 4n + \Box\, \Gamma \big)\: \fa_{n+1} \Big)  \\
&\quad\, + \Gamma_\varepsilon^n \:\log \Gamma_\varepsilon\: \big(\Box\, \fa_{n} + \mu(x)\, \fa_{n} \big) \\
&\quad\, + (n+1)\,\Gamma_\varepsilon^n \:\log \Gamma_\varepsilon\: \Big(2 \,\langle \nabla \f_, \nabla X_{n+1} \rangle + (\Box\, \f)\:X_{n+1} \Big) \\
\frak{m}_n &= (n+1) \:\Gamma_\varepsilon^n\,\Big( 2 \,\langle \nabla \Gamma, \nabla \fb_{n+1} \rangle
+ \big(4n + \Box\, \Gamma \big) \:\fb_{n+1} \Big)  \\
&\quad\, + \Gamma_\varepsilon^n\:\big(\Box\, \fb_{n} + \mu(x)\, \fb_{n} \big) \\
&\quad\, + \Gamma_\varepsilon^n\: \Big( 2 \,\langle \nabla \Gamma, \nabla \fa_{n+1} \rangle + \big(
4(2n+1) + \Box\, \Gamma \big) \: \fa_{n+1} \Big) \\
&\quad\, (n+1)\,\Gamma_\varepsilon^n\: \Big( 2 \,\langle \nabla \f_, \nabla Y_{n+1} \rangle + (\Box\, \f)\:
Y_{n+1} \Big)  \\
&\quad\, + \Gamma_\varepsilon^n\: \Big(2 \,\langle \nabla \f_, \nabla X_{n+1} \rangle + (\Box\, \f)\:X_{n+1} \Big) \\
R(x,y) &= - \frac{2}{\Gamma^3_\varepsilon}\, \big|\nabla \f \big|^2\: X_{-1} 
+\frac{1}{\Gamma^2_\varepsilon} \:|\nabla \f|^2\:\:X_0 - \frac{1}{\Gamma_\varepsilon} \:|\nabla \f|^2\:\:X_1  \\
&\quad\, - \sum_{n=0}^\infty \Gamma^n_\varepsilon
\:\log \Gamma_\varepsilon \:(n+2)(n+1)\:|\nabla \f|^2\:X_{n+2} \\
&\quad\, - \sum_{n=0}^\infty \Gamma^n_\varepsilon\: (n+2)(n+1)\:|\nabla \f|^2\:\:Y_{n+2} \\
&\quad\, - \sum_{n=0}^\infty \Gamma^n_\varepsilon\: (2n+3) \:|\nabla \f|^2\:X_{n+2}  \:.
\end{align*}
Comparing the terms~$\sim \varepsilon^2\, \Gamma^n_\varepsilon$ as described by~$R(x,y)$
with the corresponding contributions~$\sim \Gamma_\varepsilon$, one sees that these terms are of the order~\eqref{fehlerneu}.
Moreover, to this order we may replace the factors~$X_n$ by~$A_n$ and~$Y_n$
by~$B_n$. We thus obtain the transport equations of Proposition~\ref{prptransport2}.

%

\Thanks {{\em{Acknowledgments:}}
We would like to thank Claudio Dappiaggi and Maximilian Jokel for helpful discussions.
We are grateful to the referee for valuable comments on the manuscript.

\providecommand{\bysame}{\leavevmode\hbox to3em{\hrulefill}\thinspace}
\providecommand{\MR}{\relax\ifhmode\unskip\space\fi MR }
\providecommand{\MRhref}[2]{%
  \href{http://www.ams.org/mathscinet-getitem?mr=#1}{#2}
}
\providecommand{\href}[2]{#2}

\end{document}